\documentclass[moor,nonblindrev]{informs1a}   

\usepackage{dsfont,amssymb,amscd,latexsym,amsxtra,amsfonts}

\usepackage[pdfstartview=FitH, bookmarksnumbered=true,bookmarksopen=true, colorlinks=true, pdfborder=001, citecolor=blue, linkcolor=blue,urlcolor=blue]{hyperref}

\usepackage[round]{natbib}
\usepackage{bbm}
\usepackage{enumerate}
\usepackage{mathrsfs}

\usepackage{tikz}
\usetikzlibrary{calc,arrows}

\usepackage{verbatim}
\usepackage{epstopdf}
\usepackage{color}
\usepackage{bm}
\usepackage{caption}
\usepackage{subcaption}
\numberwithin{equation}{section}

\newcommand{\md}{\mathrm{d}}

\newcommand{\mR}{\mathbb{R}}

\newcommand{\mE}{\mathbb{E}}
\newcommand{\mF}{\mathbb{F}}
\newcommand{\mP}{\mathbb{P}}
\newcommand{\mQ}{\mathbb{Q}}

\newcommand{\mX}{\mathbb{X}}
\newcommand{\ind}{\mathbbm{1}}

\renewcommand{\epsilon}{\varepsilon}

\newcommand{\N}{\mathcal{N}}
\newcommand{\F}{\mathcal{F}}
\newcommand{\E}{\mathcal{E}}

\newcommand{\Si}{\mathcal{S}}

\newcommand{\Oi}{\mathcal{O}}
\newcommand{\B}{\mathcal{B}}
\newcommand{\Ui}{\mathcal{U}}

\newcommand{\cX}{\mathcal{X}}
\newcommand{\cY}{\mathcal{Y}}
\newcommand{\cZ}{\mathcal{Z}}
\newcommand{\cA}{\mathcal{A}}

\newcommand{\cG}{\mathcal{G}}
\newcommand{\cP}{\mathcal{P}}
\newcommand{\cT}{\mathcal{T}}

\newcommand{\barX}{\bar{X}}
\newcommand{\barpi}{\bar{\pi}}

\newcommand{\bP}{\mathbf{P}}
\newcommand{\bQ}{\mathbf{Q}}
\newcommand{\bF}{\mathbf{F}}

\newcommand{\br}{\mathbf{r}}
\newcommand{\blambda}{\bm{\lambda}}

\newcommand{\sF}{\mathscr{F}}

\newcommand{\bC}{\mathbf{C}}
\newcommand{\bc}{\mathbf{c}}
\newcommand{\bfb}{\mathbf{b}}

\newcommand{\mt}{\dag}
\newcommand{\defeq}{:=}
\newcommand{\eqdef}{=:}
\newcommand{\hato}{\hat{\mathrm{o}}}

\newcommand{\pp}{{\prime\prime}}

\TheoremsNumberedBySection

\def\EMAIL#1{\href{mailto:#1}{#1}}% When hyperref is used, otherwise outcomment 

\begin{document}
	
	\RUNAUTHOR{Liang, Xia and Yuan}
	
	\RUNTITLE{Dynamic portfolio selection for nonlinear law-dependent preferences}
	
	% \TITLE{Bundling Information Goods of Decreasing Value}
	\TITLE{Dynamic portfolio selection for nonlinear law-dependent preferences}
	
	\ARTICLEAUTHORS{%
		\AUTHOR{Zongxia Liang}
		\AFF{Department of Mathematical Sciences, Tsinghua University, Beijing 100084, China, \EMAIL{liangzongxia@tsinghua.edu.cn}}
		\AUTHOR{Jianming Xia}
		\AFF{RCSDS, NCMIS, Academy of Mathematics and Systems Science, Chinese Academy of Sciences, Beijing 100190, China,
		\EMAIL{xia@amss.ac.cn}}
		\AUTHOR{Fengyi Yuan}
		\AFF{Department of Mathematical Sciences, Tsinghua University, Beijing 100084, China, \EMAIL{yfy19@mails.tsinghua.edu.cn}}
		
	} % end of the block

\ABSTRACT{%
			This paper addresses the portfolio selection problem for nonlinear law-dependent preferences in continuous time, which inherently exhibit time inconsistency. Employing the method of stochastic maximum principle, we establish verification theorems for equilibrium strategies, accommodating both random market coefficients and incomplete markets.  We derive the first-order condition (FOC) for the equilibrium strategies, using a notion of functional derivatives with respect to probability distributions. Then, with the help of the FOC we obtain the equilibrium strategies in closed form for two classes of implicitly defined preferences: CRRA and CARA betweenness preferences, with deterministic market coefficients. Finally, to show applications of our theoretical results to problems with random market coefficients, we examine the weighted utility. We reveal that the equilibrium strategy can be described by a coupled system of Quadratic Backward Stochastic Differential Equations (QBSDEs). The well-posedness of this system is generally open but is established under the special structures of our problem.


	\MSCCLASS{Primary: 91B28; Secondary: 91B08 }

	\ORMSCLASS{Primary: Finance, Portfolio; Secondary: Dynamic programming/Optimal control, Models}
}%	
\maketitle
\section{Introduction.}\label{intro}

An important topic in modern financial theory is portfolio selection. Traditionally, portfolio selection problems are studied with the mean-variance (MV) criterion (\cite{Markowitz1952}) or the expected utility (EU) theory (\citet{Merton1969,Merton1971}).

The MV problems receive extensive study across various fields of finance and economics. However, it is widely recognized that the
dynamic MV problems (\citet{Li2000,Zhou2000}) exhibit time-inconsistency. To resolve the issue of time-inconsistency, the problems are approached by the intra-personal equilibrium approach, which dates back to \cite{Strotz1955}. A precise mathematical formalization of such an equilibrium in continuous time was first given by \citet{Ekeland2006} for a deterministic Ramsey model with non-exponential discounting. The equilibrium solution for continuous-time MV problems are approached by \cite{Basak2010},  \citet{BM2014}, \cite{Bjork2014}, \citet{Bjork2017}, \citet{Hu2012,Hu2017}, \cite{Dai2021} and \cite{Dai2023}.

On the other hand, EU has the advantage of  time-consistency so that dynamic programming and HJB equations can be applied (\cite{Merton1969,Merton1971}). The key property of EU that leads to its time-consistency is its linearity with respect to probability law, which is closely connected to the so-called independence axiom. However, various empirical and experimental studies show that this assumption is unrealistic, as exemplified by the famous Allais paradox (\cite{Allais1953}). This motivates emergence of various nonlinear law-dependent preferences, including the rank-dependent utility (RDU) of \citet{Quiggin1982,Quiggin1993} and the betweenness preferences of \citet{Chew1983, Chew1989} and \citet{Dekel1986}. To the best of our knowledge, only one recent paper (\cite{Hu2021}) explored equilibrium solutions to the continuous-time portfolio selection for RDU, whereas none for betweenness preferences. 

This paper aims to investigate equilibrium solutions to the continuous-time portfolio selection for a general nonlinear law-dependent preference, which 
can be represented by a functional $g: \cP_0 \to \mathbb{R}$. Here, $\cP_0$ consists of some probability measures on $\mathbb{R}$. We do not require explicit knowledge on the underlying structure of $g$ so that those implicitly-defined preference functionals $g$ can be accommodated. We also allow incomplete markets and random market coefficients. 

We utilize the open-loop formulation and the method of stochastic maximum principle to establish our verification theorems by overcoming several technical difficulties. First, to conduct variations on nonlinear functionals of probability distributions, we require a notion of derivative with respect to the distribution argument. To address this, we propose a refined version of the well-known “linear derivative" (c.f. \citet[p. 415]{Carmona2018}) that localizes the definition. We derive the FOC for the equilibrium strategies  in terms of the derivative of $g$, which brings conveniences for computations in concrete examples. Importantly, our notion of derivatives prove to be suitable for some unconventional preferences, such as those implicitly defined. Second, we encounter two distinct cases that need to be treated separately: investment strategies based on either the proportion of wealth or the dollar amount invested in the risky assets. The statements and proofs of verification theorems in these two cases exhibit significant differences. Interestingly, to the best of our knowledge, the first case (where the proportion of wealth invested is considered as the strategy) is seldom (if ever) explored in the context of time-inconsistent problems. The main technical challenge is the violation of the Lipschitz conditions because $\pi$ is not necessarily bounded. Consequently, usual estimates in the theory of stochastic differential equations (SDE) are no longer applicable. To tackle this issue, we employ the theory of BMO (bounded mean oscillation) martingales, along with estimates of exponential processes; see the proof of Theorem \ref{verification-pc} in Appendix \ref{proof:verification-pc}. 

As applications of our verification theorems, we explore three types of preferences, which are new to the continuous-time finance literature. 
\begin{itemize}
 \item[(1)] Assume that $g$ is implicitly defined by \[
 \mE[F(X/g(\mP_X))]=0,
 \]
where $F:(0,\infty)\to\mR$ is a concave and increasing function with $F(1)=0$.  
 Such a preference is called a CRRA (constant-relative-risk-aversion) betweenness preference; see \citet[Section 25.2]{Back2017}. 
 We use one of our verification theorems to obtain the equilibrium strategies in closed form for CRRA betweenness preferences when the market coefficients are deterministic.
  \item[(2)] Similarly, a CARA (constant-absolute-risk-aversion) betweenness preference functional $g$ is implicitly given by
 \[
 \mE [F(X-g(\mP_X))]=0,
 \]
 where $F:\mR\to\mR$ is a concave and increasing function with $F(0)=0$.   
 We also provide the equilibrium strategies in closed form for such a preference when the market coefficients are deterministic.
 \item[(3)] To illustrate the application of our verification theorems to the case of random market coefficients, we consider a weighted utility given by
 \[
 g(\mP_X)\defeq \frac{\mE \left[X^{1-\rho}X^{\gamma}\right]}{(1-\rho)\mE\left[ X^{\gamma}\right]},
 \]
 where $-1<\gamma\leq 0$, $\gamma\leq \rho<\gamma+1$ (c.f. \citet[Exercise 25.4]{Back2017}). We show that the equilibirum solutions are now described by a system of two-dimensional QBSDEs. It is acknowledged that there is no general well-posedness result for multi-dimensional QBSDEs. We prove the well-posedness of the QBSDEs in our example, relying on the “pure linear-quadratic" structure of the generator and the equivalence of BMO norms under the change of measures. For details and literature review, see Section {\ref{WUrandom}}.
\end{itemize}

To sum up, the contributions of our paper are threefold. First, we establish a general framework for the study of dynamic portfolio selection problems with nonlinear preferences, which applies to a broad class of preferences, including implicit-defined preferences. Second, we establish a verification theorem when the risky investment proportion is regarded as the strategy. The BMO martingale theory and the estimates of exponential processes are used to tackle the technical difficulties. Finally, we investigate concrete examples with closed-form solutions when the market coefficients are deterministic. We also investigate an example with random market coefficient and show that a two-dimensional QBSDE can be used to determine the equilibrium strategies. Existing results on the well-posednness of multi-dimensional QBSDEs are not applicable to our case. Thus, we prove the well-posedness by making full use of the special structure of our problem. 

The rest of the paper is organized as follows. Section \ref{formulation} formulates the problem.
Section \ref{verificationthm} provides the verification theorems.  
Sections \ref{generalCRRA}--\ref{WUrandom} study several concrete examples in details. Section \ref{conclu} concludes the paper. The appendix includes technical lemmas and proofs. We also recover some existing results by our approach in the Online Companion.

\section{Problem formulation.}\label{formulation}

\subsection{Market model.}
\label{admissiblestrategy}
{Let $(\Omega,\F,\mF,\mP)$ be a filtered probability space, where  $\mF=\{\F_t\}_{0\le t\le T}$ is the $\mP$-augmentation of the natural filtration generated by a $(d+k)$-dimensional standard Brownian motion $W=(W^{\Si},W^{\Oi})$, constant $T>0$ is the time horizon and $\F=\F_T$.}  Here, the superscripts $\Si$ means that the $d$-dimensional Brownian motion $W^{\Si}$ drives the stock prices, and $\Oi$ means “orthogonal”, indicating that the $k$-dimensional Brownian motion $W^{\Oi}$ models orthogonal risks. We assume that there are $d$-stocks in the market, and their price processes $\{S^i_t,i=1,2,\cdots,d,t\in [0,T]\}$ follow the dynamics
\begin{equation}\label{stockprice}
	\left\{
	\begin{aligned}
	&\md S^i_t = S^i_t[\theta^i(t)\md t+\sigma^i(t)\cdot \md W^{\Si}_t],\\
	&S^i_0=s^i_0>0,
	\end{aligned}
\right.
\end{equation}
where the market coefficients $\theta$ and $\sigma$ as well as the interest rate $r$ always satisfy the following standing assumption.

\begin{assumption}\label{ass:thetasigma}
The interest rate $r=0$. 
The processes $\theta:[0,T]\times \Omega \to \mR^d$ and $\sigma:[0,T]\times \Omega \to \mR^{d\times d}$ are $\mF$-predictable.  The process $\theta$ is bounded. There exist some strictly positive constants $c_1$ and $c_2$ such that
$c_1|a|^2\le |\sigma(t)^\mt a|^2\le c_2|a|^2$ a.s. for every $a\in\mR^d$ and almost every $t\in[0,T]$. (Hereafter, we denote by $M^\mt$ the transpose of a matrix $M$.)
\end{assumption}

We emphasize that the market coefficients are adapted to the whole filtration, generated by overall risk $W$. Let
\begin{equation}\label{eq:kappa}
\kappa(t)=(\sigma(t))^{-1}\theta(t), \quad t\in[0,T].
\end{equation}
Under Assumption \ref{ass:thetasigma}, the process $\kappa$ is bounded. 
\vskip 5pt
A trading strategy is an $\mF$-predictable process $\{\pi_s,0\leq s\leq T\}$ such that 
\[
\int_0^T (|\pi_s^\mt\theta(s)|+|\pi_s^\mt\sigma(s)|^2)\md s<\infty,\ \text{ $\mP$-a.s.}
\]
When $\pi$ models the proportion of the wealth invested into the stocks,  the self-financing wealth process $\{X^{\pi}_t,0\leq t\leq T\}$  satisfies the following SDE
\begin{equation}\label{wealthdynamic:proportion}
	\left\{
	\begin{aligned}
	&\md X^\pi_t=X^\pi_t\pi_t^\mt\theta(t)\md t+X^\pi_t\pi_t^\mt \sigma(t)\cdot\md W^{\Si}_t,\\	
	&X^\pi_0=x_0,
    \end{aligned}
\right.
\end{equation}
where $x_0$ is the initial wealth.\\
When $\pi$ models the dollar amount invested into the stocks,  the self-financing wealth process $\{X^{\pi}_t,0\leq t\leq T\}$  satisfies the following SDE
\begin{equation}\label{wealthdynamic:amount}
	\left\{
	\begin{aligned}
	&\md X^\pi_t=\pi_t^\mt\theta(t)\md t+\pi_t^\mt \sigma(t)\cdot\md W^{\Si}_t,\\	
	&X^\pi_0=x_0.
    \end{aligned}
\right.
\end{equation}

\subsection{Equilibrium strategy.}
Hereafter, for any $\F_T$-measurable random variable $\xi$ and $t\in [0,T)$, $\mP^t_\xi$ is the regular conditional law of $\xi$ given $\F_t$ under probability $\mP$. %$\cL(X|\F_t)$ . 
In particular, $\mP_\xi=\mP^0_\xi$ is the law of $\xi$ under probability $\mP$. Throughout this paper,  
$\mX$ is a fixed nonempty interval and $\cP(\mX)$ denotes the set of all probability measures on $\mX$.
In our examples, we usually take $\mX=\mR$ or $\mX=(0,\infty)$.
$\cP_0$ is a nonempty convex subset of $\cP(\mX)$.

\begin{definition}\label{adm-strategy}
A trading strategy $\pi$ is called admissible if $\mP^t_{X^\pi_T}\in \cP_0$ a.s. for each $t\in [0,T)$.
\end{definition}

The set of all admissible strategies is denoted by $\cA$. 

We assume that,  at each time $t\in[0,T)$, the investor’s preference for the terminal wealth $X^\pi_T$ can be represented by 
$g(\mP^t_{X^\pi_T})$, where  
$$g:\cP_0\to \mR$$
is a functional. 
Represented by $g$, the investor’s preference at each time $t\in[0,T)$ for the terminal wealth $X^\pi_T$ depends only on the condition law $\mP^t_{X^\pi_T}$.  Law-dependent preferences include as special cases the well-known mean-variance utility and rank-dependent utility of \citet{Quiggin1982,Quiggin1993}, which are all discussed in the Online Companion, as well as the betweenness preferences of \citet{Chew1983, Chew1989} and \citet{Dekel1986}. In this paper, we will first study the case of general law-dependent preferences in Section \ref{verificationthm} and then focus on the special cases of the CRRA and CARA betweenness preferences and the weighted utility preference in Sections \ref{generalCRRA}--\ref{WUrandom}.

For each $t\in[0,T]$, $p\in[1,\infty]$ and $m\ge 1$,  we use  $L^p(\F_t,\mR^m)$ to denote the set of all $L^p$-integrable, $\mR^m$-valued and $\F_t$-measurable random variables. For simplicity, we write $L^{p}(\F_t)$ for $L^p(\F_t,\mR)$.
Throughout this paper,  $\barpi\in \cA$ is fixed.  For any $t\in[0,T)$, $\epsilon\in(0,T-t)$, and $\varphi\in L^\infty(\F_t,\mR^d)$, the perturbed strategy  $\barpi^{t,\epsilon,\varphi}$ is given by $\barpi^{t,\epsilon,\varphi}\defeq\barpi+\varphi\ind_{[t,t+\epsilon)}$, i.e.,
	\[\barpi^{t,\epsilon,\varphi}_s=
	\left\{
	\begin{aligned}
		&\barpi_s+\varphi, &s\in [t,t+\epsilon),\\
		&\barpi_s, &s\notin [t,t+\epsilon).
	\end{aligned}
\right.
	\]
We will write $\barX\defeq X^{\barpi}$ and $\barX^{t,\epsilon,\varphi}\defeq X^{\barpi^{t,\epsilon,\varphi}}$ for simplicity.

We will consider two types of equilibrium strategies which are defined as follows.

\begin{definition}\label{adm-equi}
	\begin{description}
	
	\item[(a)] $\barpi$ is called a {\bf Type-I} equilibrium strategy, if for any $t\in[0,T)$ and $\varphi\in L^\infty(\F_t,\mR^d)$ such that $\barpi^{t,\epsilon,\varphi}\in\cA$ for all sufficiently small $\epsilon>0$, we have
	\[
	\limsup_{\epsilon\to 0 }\frac{1}{\epsilon}\left(g\left(\mP^{t}_{\barX^{t,\epsilon,\varphi}_T}\right)-g\left(\mP^{t}_{\barX_T}\right)\right) \leq 0.
	\]
		\item[(b)]$\barpi$ is called a {\bf Type-II} equilibrium strategy if, for a.e. $t\in [0,T)$, any $\varphi\in L^\infty(\F_t,\mR^d)$ such that $\barpi^{t,\epsilon,\varphi}\in\cA$ for all sufficiently small $\epsilon>0$, and any $\zeta\in L^{\infty}(\F_t)$ with $\zeta\geq 0$, we have  
	\[
	\limsup_{\epsilon\to 0}\mE\left[\frac{1}{\epsilon}\left({g\left(\mP^{t}_{\barX^{t,\epsilon,\varphi}_T}\right)-g\left(\mP^{t}_{\barX_T}\right)}\right)\zeta\right]\leq 0.
	\] 
	\end{description}
\end{definition}

\begin{remark}
The definition of type-I equilibrium is frequently used in time-inconsistent control literature; see e.g. \citet{Bjork2017} and \cite{Hamaguchi2021}. On the other hand, the definition of type-II equilibrium is recently proposed in \citet{Hamaguchi2021a}. For relations of these two types of equilibrium, see Remark 4.2 of \cite{Hamaguchi2021a}. We think type-II equilibrium is more appropriate to deal with random market coefficients which bring regularity issues; see Remark \ref{regularityissue} for detailed discussions.
\end{remark}

We will characterize these two types of equilibria in Section \ref{verificationthm}.

\subsection{Notations and Preliminaries.}\label{notations}
For the readers' convenience, we list here the notations that are frequently used throughout this paper. 
\begin{itemize}
\item $\langle f,\mu\rangle:=\int_\mX f(x)\mu(dx)$ for a Borel function $f:\mX\to\mR$ and a signed measure $\mu$ on $\mX$.
\item For any $\mu\in\cP(\mX)$, $L^1_\mu$ denotes the set of all Borel functions $f:\mX\to\mR$ such that 
$$\int_\mX |f(x)|\mu(dx)<\infty.$$
		\item $L^0(\mF,\mR^m)$: the space of $\mF$-progressively measurable processes taking values in $\mR^m$.
	\item $L^{p,q}(\mF,\mR^m)$, $1\leq p,q<\infty$: the space of processes $Y\in L^0(\mF,\mR^m)$ such that
	\[
	\|Y\|_{L^{p,2}(\mF,\mR^m)}\defeq \left(\mE\left[\left(\int_0^T |Y_s|^q\md s  \right)^{p/q}\right]\right)^{1/p}<\infty.
	\]
	When $p=q$, we write $L^p(\mF,\mR^m)=L^{p,p}(\mF,\mR^m)$.
	\item $L^\infty(\mF,\mR^m)$: the space of bounded processes $Y\in L^0(\mF,\mR^m)$. 
	\item $S^p(\mF,\mR^m)$, $1\leq p<\infty$: the space of processes $Y\in L^0(\mF,\mR^m)$ with right continuous paths such that
	\[
	\|Y\|_{S^p(\mF,\mR^m)}\defeq \left(\mE\left[\sup_{s\in [0,T]}|Y_s|^p\right]\right)^{1/p}<\infty.
	\]
	\item For a probability measure $\mP'$ and $t\in[0,T]$, we write $\mE^{\mP’}_t[\cdot]$ for $\mE^{\mP'}[\cdot|\F_t]$. When $\mP'=\mP$, the superscript is omitted. 
	\item {$\cT_{[s,t]}$: the set of all $\mF$-stopping times $\tau$ such that  $s\leq \tau\leq t$ a.s., $0\le s\le t\le T$.
	\item For a probability $\mP'$, let $H^{d}_{{\rm BMO}(\mP')}=\{\pi\in L^0(\mF,\mR^d):\|\pi\|_{{\rm BMO}(\mP')}<\infty\}$, where
		\[
		\|\pi\|^2_{{\rm BMO}(\mP’)}\defeq \sup_{\tau\in\cT_{[0,T]}} \left\| \mE^{\mP’}_{\tau}\left[\int_\tau^T |\pi_s|^2\md s \right] \right\|_\infty.
		\] }
		We write $H^d_{\rm BMO}=H^d_{{\rm BMO}(\mP)}$, $H_{\rm BMO}=H^{1}_{\rm BMO}$ and $H_{{\rm BMO}(\mP')}=H^1_{{\rm BMO}(\mP')}$.
    \item $\cZ^W(\alpha)=\{\cZ^W_s(\alpha): 0\leq s\leq T\}$: the exponential martingale related to the process $\alpha$ and Brownian motion $W$, as defined by $\cZ^W_s(\alpha)\defeq \exp(\int_0^s \alpha_r\cdot\md W_r-\frac{1}{2}\int_0^s\|\alpha_r\|^2\md r)$.
	\item Let $p>1$ and $\eta\in L^p(\F_T)$.  The process $Z^\eta=(Z^{\eta,\Si},Z^{\eta,\Oi})$ is defined as in the following lemma, which is not original but provided here for the readers’ convenience. 
	\end{itemize}

\begin{lemma}\label{MRM-p}
	Let $p>1$ and $\eta\in L^p(\F_T)$. Let $Y_s\defeq \mE_s[\eta]$, $s\in [0,T]$. Then $Y:=\{Y_s,s\in[0,T]\}\in S^p(\mF,\mR)$. Moreover, there exists a unique pair $(Z^{\eta,\Si}, Z^{\eta,\Oi})\in L^{p,2}(\mF,\mR^{d})\times L^{p,2}(\mF,\mR^{k})$ such that
	\begin{equation}\label{MRM-eq}
	Y_s = \mE[\eta]+\int_0^s Z_r^{\eta,\Si}\cdot \md W_r^\Si+\int_0^s Z_r^{\eta,\Oi}\cdot \md W_r^\Oi,\quad s\in[0,T].
	\end{equation}
\end{lemma}
\proof{Proof.}
	It is clear that $Y$ is an $L^p$-martingale. The martingale representation theorem implies that
	there is a unique pair $(Z^{\eta,\Si}, Z^{\eta,\Oi})\in L^0(\mF,\mR^{d})\times L^0(\mF,\mR^{k})$ such that (\ref{MRM-eq}) holds. Noting that $|Y|$ is a nonnegative submartingale and using Doob's maximal inequality, we have
	\[
	\mE\left[\sup_{s\in[0,T]}|Y_s|^p\right]\leq \left(\frac{p}{p-1}\right)^p \mE|Y_T|^p<\infty,
	\]
	i.e., $Y\in S^p(\mF,\mR)$. Applying the Burkholder-Davis-Gundy inequality and using (\ref{MRM-eq}), we know $(Z^{\eta,\Si}, Z^{\eta,\Oi})\in L^{p,2}(\mF,\mR^{d})\times L^{p,2}(\mF,\mR^{k})$.
\hfill \Halmos\endproof

\section{The verification theorems.}\label{verificationthm}

In this section we establish some verification theorems. To this end, we first propose a notion of differentiability of a law-dependent preference functional. Our notion is in spirit in line with but weaker than the so-called “linear derivative”( c.f. \citet[p. 415]{Carmona2018}). We will use this notion to deal with the implicitly-defined preference functionals. 
\begin{definition}\label{derivativedef}
	Let $\mu,\nu\in\cP_0$. Let $[\nu,\mu]:=\{s\nu+(1-s)\mu: s\in[0,1]\}$. We say that $g$ is differentiable on $[\nu,\mu]$ if the function $[0,1]\ni s \mapsto g(s\nu+(1-s)\mu)$ is almost everywhere differentiable and there exists a function $\nabla g: \cP_0\times \mX\to\mR$ such that, for almost every $s\in [0,1]$, 
	$$\nabla g(s\nu+(1-s)\mu,\cdot)\in L^1_\nu\cap L^1_\mu\text{ and }\frac{\md}{\md s}g(s\nu+(1-s)\mu)=\langle \nabla g(s\nu+(1-s)\mu,\cdot) ,\nu-\mu\rangle.$$
\end{definition}

We impose the following standing assumption on $g$.

\begin{assumption}\label{gassumption1} \
	\begin{description}
		\item[(a)]$g$ is differentiable on $[\nu,\mu]$ for any $\mu,\nu\in\cP_0$, and the derivative $\nabla g:\cP_0\times \mX\to \mR$ satisfies that $\int_\mX|\nabla g(\mu,x)|\nu(\md x)<\infty$ for any $\mu,\nu\in\cP_0$.
		\item[(b)] $\nabla g(\mu,\cdot)$ is concave for any $\mu\in \cP_0$.
		\item[(c)] There exist functions $M_0:\cP_0\times \cP_0\to \mR$ and $M_1:\cP_0\times \cP_0\to \mR_+$ such that 
		\begin{equation}\label{quasiconcave}
		g(\mu_1)-g(\mu_0)\leq M_1(\mu_1,\mu_0)\langle \nabla g(\mu_0,\cdot),\mu_1-\mu_0\rangle+M_0(\mu_1,\mu_0),\quad \forall \mu_0,\mu_1\in\cP_0.	
		\end{equation} 
		\end{description}
\end{assumption}

\begin{remark}
We make a few comments on Assumption \ref{gassumption1}. First, if $g(\mu)=\langle U,\mu\rangle$ (i.e., the EU with a utility function $U$), then by Definition \ref{derivativedef}, $\nabla g(\mu,\cdot)=U$. Therefore, Assumption \ref{gassumption1}(b) is a natural generalization of the concavity of utility function to our context. Second, Assumption \ref{gassumption1}(c) is a relaxation of the concavity of $g$, which holds automatically in the EU case because $g$ is linear.
\end{remark}

By Assumption \ref{gassumption1}(b), there exists a super-gradient map of $\nabla g(\mu,\cdot)$ for fixed $\mu$, which is a {\it set-valued} map, denoted by $\partial_x \nabla g(\mu,\cdot):\mX\mapsto 2^\mR$. If $\nabla g (\mu,\cdot)$ is smooth, we abuse $\partial_x \nabla g(\mu,\cdot)$ to denote its derivative with respect to $x$. 

We are now ready to provide our verification theorems. For clarity we state and prove the theorems in two separated cases: proportion strategies and dollar-amount strategies. 

\subsection{Proportion strategies.}\label{subsectionveripc}

This subsection provides a verification theorem when the control variable $\pi$ is regarded as the proportion of wealth invested into the risky assets.

\begin{theorem}\label{verification-pc}
	Fix a $\barpi \in H^d_{\rm BMO}\cap \cA$. Consider the following conditions:\footnote{We emphasize that the superscript $t$ of $\xi^t$ denotes the dependence on $t$ rather than the power of $t$.}
	\begin{description}
	\item[(a)] There exists $p>1$ such that for any $t\in [0,T)$, there exists an $\F_T$-measurable $\xi^t$ such that $\xi^t\in \partial_x \nabla g(\mP^{t}_{\barX_T},\barX_T)$ almost surely and $\barX_T \xi^t\in L^p(\F_T)$. Moreover, the following holds almost surely:
	\begin{equation}\label{FOC}
		(\kappa(t)-\sigma^\mt (t)\barpi_t)\mE_t[\barX_T\xi^t]+Z^{\barX_T\xi^t,\Si}(t)=0, \quad t\in [0,T].
	\end{equation}
	\item[(b)] For any $t\in [0,T)$, $Z^{\barX_T \xi^t,\Si}\in S^{p}(\mF,\mR^d)$. 
	\item[(b')] $(\kappa(s)-\sigma^\mt (s)\barpi_s)\mE_s[\barX_T\xi^t]+Z^{\barX_T\xi^t,\Si}(s)=D(t)\sum_{i=1}^M A_i(s)B_i(s)C_i(t)$ for some $M\geq 1$, where $B_i,C_i,D$ have right continuous paths, $D>0$ and $A_i\in H^d_{\rm BMO}$. Moreover, for a.e. $t\in [0,T)$, there exist $m>1$ and some constant $K> 0$ such that, for any sufficiently small $\epsilon>0$, we have
\begin{equation}\label{1/2Holder}
	\sup_{i=1,2,\cdots,M}\mE\left[\sup_{t\leq s\leq t+\epsilon}(D(t)|B_i(s)||C_i(t)-C_i(s)|)^m\right]\leq K \epsilon^{m/2}.
\end{equation}
\item[(c)] For any $t\in [0,T)$ and $\varphi\in L^{\infty}(\F_t,\mR^d)$, we have 
\begin{align}
	&\limsup_{\epsilon\to 0} |M_1(\mP^{t}_{\barX^{t,\epsilon,\varphi}_T},\mP^{t}_{\barX_T})|<\infty,\label{Mbound}\\
	&\lim_{\epsilon\to 0}\frac{1}{\epsilon}|M_0(\mP^{t}_{\barX^{t,\epsilon,\varphi}_T},\mP^{t}_{\barX_T})|=0,\label{Nbound}
\end{align}
almost surely.
\item[(c')] For a.e. $t\in [0,T)$, any $\varphi\in L^{\infty}(\F_t,\mR^d)$, and any $n\geq 1$, we have 
\begin{align}
	&\limsup_{\epsilon\to 0} \mE\left[|M_1(\mP^{t}_{\barX^{t,\epsilon,\varphi}_T},\mP^{t}_{\barX_T})|^n\right]<\infty,\label{Mintegral}\\
	&\lim_{\epsilon\to 0}\frac{1}{\epsilon}\mE\left[|M_0(\mP^{t}_{\barX^{t,\epsilon,\varphi}_T},\mP^{t}_{\barX_T})|\right]=0.\label{Nintegral}
\end{align}
		\end{description}	
Then, $\barpi$ is a Type-I equilibrium proportion strategy if we have (a), (b) and (c), and is a Type-II equilibrium proportion strategy if we have (a), (b') and (c').
\end{theorem}
\proof{Proof.}
	See Appendix \ref{proof:verification-pc}.
\hfill \Halmos\endproof 

\vskip 5pt
Condition (\ref{FOC}) can be viewed as the FOC or “equilibrium condition" that determines the candidate equilibrium strategies; see Sections \ref{generalCRRA} and \ref{WUrandom} for applications.  

\vskip 5pt

\begin{remark}\label{regularityissue}
	Let us make comments on Conditions (b) and (b'). As is well-known in the time-inconsistent control literature, the {\it diagnol} processes of certain two-parameter random fields are crucial for the analysis. Indeed, in our equilibrium condition (\ref{FOC}), there appears the diagnol processes. However, as pointed out by \cite{Hamaguchi2021a}, the diagnol process $t\mapsto Z(t,t)\in \mR^k$ is not even well-defined if $Z(t,\cdot)$ is just in $L^{p,2}(\mF,\mR^k)$; see Remark 2.3 and Example 2.4 therein. In the present paper, without further assumptions, we only have $\barX q^{t,\Si}\in L^{p_1,2}(\mF,\mR^d)$, thus the well-posedness of the equation (\ref{FOC}) is a serious issue. With (b) or (b'), it is apparent that the diagnol process $t\mapsto \barX_t q^{t,\Si}(t)$ is well-defined. Moreover, we know that this diagnol process belongs to the space $L^1(\mF,\mR^d)$; see also discussions on the diagnol process in \cite{Hernandez2023}. This is the main reason we choose to impose such conditions. The reason why we impose two distinctive conditions is that they are applicable to different examples, and are exclusive to each other. We also remark that because of the characteristics of our problems, we do not expect the further regular assumptions imposed in \citet[Lemma 2.7]{Hamaguchi2021a} to hold. Indeed, in our weighted utility example (see Section \ref{WUrandom} for details), $\barX_sq^{t,\Si}(s)=c_1\tilde{Z}_2(s)Y_2(s)/Y_1(t)+c_2\tilde{Z}_1(s)Y_1(s)Y_2(t)/Y_1(t)^2$, where $\tilde{Z}\in H^d_{\rm BMO}$. Therefore, measuring in any $L^m$-norm, the regularity with respect to $t$ is $1/2$-H$\ddot{\mathrm{o}}$lder, hence is not absolutely continuous or $C^1$;  see (\ref{1/2Holder}) of Condition (b').
\end{remark}

\subsection{Dollar-amount strategies.}\label{subsectionverificationnc}
This subsection provides verification theorem when the control variable $\pi$ is regarded as the dollar amount, instead of the proportion of wealth, invested into the risky assets. 

To proceed, we first introduce a fact as follows.  { Let $p>1$. For any $\xi^t\in L^p(\F_T)$, by Lemma \ref{MRM-eq}, {we have $\{\mE_s[\xi^t],s\in [0,T]\}\in S^p(\mF;\mR)$ and $Z^{\xi^t}\in L^{p,2}(\mF,\mR^{d+k})$}. }

We now have the following version of verification theorem.
\begin{theorem}\label{verification-nc}
	$\barpi \in L^2(\mF,\mR^d)\cap\cA$ is a Type-I equilibrium dollar-amount strategy if the following conditions are satisfied:	\begin{description}
		\item[(a)]There exists $p>1$ such that for any $t\in [0,T)$, there exists a $\xi^t\in L^p(\F_T)$ with $\xi^t\in \partial_x \nabla g(\mP^{t}_{\barX_T},\barX_T)$ almost surely and 
		the following holds almost surely:
		\begin{equation}\label{FOCnc}
		\kappa(t)\mE_t[\xi^t]+Z^{\xi^t,\Si}(t)=0,\quad\forall t\in [0,T].
		\end{equation}
		\item[(b)]For any $t\in [0,T)$, $Z^{\xi^t,\Si}\in S^{p}(\mF,\mR^d)$.
		\item[(c)] For any $t\in [0,T)$ and $\varphi\in L^{\infty}(\F_t,\mR^d)$, we have (\ref{Mbound}) and (\ref{Nbound}).
	\end{description}	
\end{theorem}
\proof{Proof.}
	See Appendix \ref{proof:verification-nc}.
\hfill \Halmos\endproof

 \vskip 5pt
 Theorem \ref{verification-nc} will be applied to investigate the equilibrium strategies for the CARA betweenness preferences, the mean-variance preferences and the rank dependent utilities; see Section \ref{generalCARA} and the Online Companion. 

\section{CRRA betweenness preferences.}\label{generalCRRA}
In this section, we apply Theorem \ref{verification-pc} to investigate a large class of betweenness preferences: the CRRA betweenness preferences. We first derive some useful properties of the preference functional $g$, including the expression of its derivative. Then, assuming the market coefficients are deterministic, we make an ansatz and transform the equilibrium condition \eqref{FOC} into an ODE, which can be solved in closed-form under certain technical conditions. Finally, we verify that the solution of the ODE gives a type-I equilibrium based on Theorem \ref{verification-pc}.

\subsection{Some properties of the preferences.}\label{CRRABP}
We take $\mX=(0,\infty)$ and the preference functional $g$ is implicitly defined by
\begin{equation}\label{gCRRA}
	\mE \left[F\left( \frac{X}{g(\mP_X)}  \right)\right]=0,
\end{equation}
where $F$ is a function satisfying the following assumption.
\begin{assumption}\label{ass:F:crra}
$F:(0,\infty)\to \mR$ is twice continuously differentiable, $F(1)=0$, $F^\prime(x)>0$ and $F^\pp(x)<0$ for all $x\in(0,\infty)$. Moreover, there exist $C>0$ and $\alpha_0>0$ such that 
\[
|F'(x)|+|F''(x)|\leq C(x^{-\alpha_0}+x^{\alpha_0}),\quad\forall x>0.
\]
\end{assumption}

In this section, the set $\cP_0$  
satisfies the following assumption.
\begin{assumption}\label{ass:P0}
$\cP_0$ is a nonempty convex subset of $\cP((0,\infty))$ satisfying the following two conditions:
\begin{description}
\item[(a)] If $\mP_X\in\cP_0$, then $\mP_{\lambda X}\in\cP_0$ for all $\lambda>0$.
\item[(b)] $\int_0^\infty (|F(x)|+xF^\prime(x)-x^2 F^\pp(x))\mu(dx)<\infty$ \ for all $\mu\in\cP_0$.
 \end{description}
\end{assumption}
 
The preferences given by (\ref{gCRRA}) have both the  betweenness property  of \citet{Chew1983, Chew1989} and \cite{Dekel1986} and the positive homogeneity ($g(\mP_{\lambda X})=\lambda g(\mP_X)$ if $\lambda>0$ and $\mP_X\in\cP_0$). Therefore, they are called the CRRA betweenness preferences (c.f. \cite{Back2017}). 
It should be noted that $g(\mP_X)$ is the certainty equivalent of $X$ for the CRRA betweenness preference.\footnote{Because $g(\delta_x)=x$ for any $x>0$ and the Dirac measure $\delta_x$, we have $g\left(\delta_{g(\mP_X)}\right)=g(\mP_X)$, implying that $g(\mP_X)$ is the certainty equivalent of $X$.} In particular, if $F=U_\gamma$ for some CRRA utility function\footnote{\label{footnotecrra}Keep in mind the difference between the CRRA utility functions and the CRRA betweenness preferences: a CRRA utility function is a function $U_\gamma$ given by \eqref{crrau}, whereas a CRRA betweenness preference is represented by a functional $g:\cP_0\to\mR$ satisfying \eqref{gCRRA}.} 
\begin{equation}\label{crrau}
 U_\gamma(x)=
\begin{cases}
	\frac{x^\gamma-1}{\gamma}, &\gamma\neq 0,\,\gamma<1,\\
	\log x, &\gamma =0,
\end{cases}
\end{equation}
then $g(\mP_X)$ coincides the certainty equivalent $U_\gamma^{-1}(\mE[U_\gamma(X)])$ of $X$ for the von Neumann and Morgenstern expected utility. In general, the two certainty equivalents are not necessarily equal. 

We have the following result about the derivative of $g$.
\begin{lemma}\label{CRRAderivativelemma}
	Let $g$ be given by (\ref{gCRRA}). Then, under Assumptions \ref{ass:F:crra}--\ref{ass:P0}, we have, for any $\mu\in\cP_0$ and $x>0$, 
	\begin{equation}\label{derivativeCRRA}
		\nabla g(\mu,x)=F\left(\frac{x}{g(\mu)}\right)\frac{g(\mu)^2}{\int_0^\infty yF'\left(\frac{y}{g(\mu)}\right)\mu(\md y) }.
	\end{equation}
	Moreover, Assumption \ref{gassumption1} is satisfied. In particular, Assumption \ref{gassumption1}(c) is satisfied with the choice $M_0=0$ and
	\begin{equation}\label{WUmultiM1}
		M_1(\mu_1,\mu_0)=\frac{g(\mu_1)}{g(\mu_0)}\frac{\int_0^\infty xF'\left(\frac{x}{g(\mu_0)}  \right)\mu_0(\md x)}{\int_0^\infty xF'\left(\frac{x}{g(\mu_0)}  \right)\mu_1(\md x)}.
	\end{equation}
\end{lemma}
\proof{Proof.}
	Let $\mu_0,\mu_1\in\cP_0$ and $\mu_s\defeq s\mu_1+(1-s)\mu_0$, $s\in[0,1]$. Then  we have
	\[
	s\int_0^\infty F\left(\frac{x}{g(\mu_s)}  \right) \mu_1(\md x)+(1-s)\int_0^\infty F\left(\frac{x}{g(\mu_s)}  \right) \mu_0(\md x)=0.
	\]
	Taking $\frac{\md }{\md s}$ on both sides, we get
	\[
	\frac{\md }{\md s}g(s\mu_1+(1-s)\mu_0)=\int_0^\infty    \frac{g(\mu_s)^2F\left( \frac{x}{g(\mu_s)}  \right)}{\int_0^\infty yF'\left(\frac{y}{g(\mu_s)}\right)\mu_s(\md y)}  (\mu_1-\mu_0)(\md x),
	\]
	which yields \eqref{derivativeCRRA} by Definition \ref{derivativedef}.
Obviously, Assumption \ref{gassumption1}(a) is implied by \eqref{derivativeCRRA} and Assumption \ref{ass:P0}.
 Assumption \ref{gassumption1}(b) is clearly implied by the concavity of $F$. 

 \vskip 5pt
 Finally, we  show that $g$ satisfies Assumption \ref{gassumption1}(c). Using (\ref{gCRRA}) and the concavity of $F$, we have
	\begin{align*}
		\int_0^\infty F\left(\frac{x}{g(\mu_0)}\right)(\mu_1-\mu_0)(\md x)&=\int_0^\infty \left(F\left(\frac{x}{g(\mu_0)}\right)-F\left(\frac{x}{g(\mu_1)}\right)     \right)\mu_1(\md x)\\
		&\geq \int_0^\infty xF'\left(\frac{x}{g(\mu_0)}   \right)\frac{g(\mu_1)-g(\mu_0)}{g(\mu_1)g(\mu_0)}\mu_1(\md x).
	\end{align*} 
	Thus 
	\[
	g(\mu_1)-g(\mu_0)\leq \int_0^\infty \frac{F\left( \frac{x}{g(\mu_0)}\right)g(\mu_1)g(\mu_0)  }{\int_0^\infty yF'\left( \frac{y}{g(\mu_0)} \right)\mu_1(\md y)}(\mu_1-\mu_0)(\md x).
	\]
	Combing this with (\ref{derivativeCRRA}), we know that Assumption \ref{gassumption1}(c) holds with $M_1$ given by (\ref{WUmultiM1})\ \  and $M_0\equiv 0$.\hfill \Halmos
\endproof

\subsection{Three important functions.}

We now introduce three important functions that will be used. 

The first function $H:[0,\infty)\to(0,\infty)$ is related to the certainty equivalent of log-normal distributions. 
Consider a random variable $\xi\sim\N(0,1)$. Hereafter, $\N(0,1)$ denotes the standard normal distribution. Lemma \ref{Hdeflemma} below shows that, under the following assumption, a function $H$ is well defined by 
 \begin{equation}\label{Hdef}
 \mE\left[F\left(e^{\sqrt{y}\xi}/H(y)\right)\right]=0,\quad y\ge0.
 \end{equation}  
 
\begin{assumption}\label{ass:ui:F}
For any $y_0\in(0,\infty)$ and $z_0\in(0,1)$, the family 
$$\left\{F\left({1\over z}e^{\sqrt{y}\xi}\right): y\in[0,y_0],\, z\in[z_0,1/z_0]\right\}$$
is uniformly integrable, where $\xi\sim\N(0,1)$. 
\end{assumption}

\begin{lemma}\label{Hdeflemma}
Let $g$ be given by (\ref{gCRRA}). Then, under Assumptions \ref{ass:F:crra}--\ref{ass:P0} and \ref{ass:ui:F}, $H$ is well defined,  $H$ is continuous on $[0,\infty)$ and $H(0)=1$.
\end{lemma}

\proof{Proof.} Let 
$$f(y,z)=\mE\left[F\left({1\over z}e^{\sqrt{y}\xi}\right)\right],\quad y\ge 0, z>0.$$
It is easy to see that $f$ is continuous on $[0,\infty)\times(0,\infty)$ and, for each $y\ge0$, $f(y,z)$ is strictly decreasing w.r.t. $z$. Moreover, the Monotone convergence theorem implies that, for each $y\ge0$, 
$\lim_{z\to\infty}f(y,z)=F(0+)<0$ and $\lim_{z\to 0}f(y,z)=F(\infty)>0$.
Thus the conclusion follows.
\hfill \Halmos\endproof 

 \vskip 5pt
By definition,
$$H(y)=g(\mP_{e^{\sqrt{y}\xi}}),\quad y\ge0.$$ 

The second function $G:[0,\infty)\to(0,\infty)$ is defined by
\begin{equation}\label{Gdef}
G(y)\defeq \frac{H(y)\cdot \mE\left[e^{\sqrt{y}\xi}F’\big(e^{\sqrt{y}\xi}/H(y)\big)\right]}
{-\mE\left[e^{2\sqrt{y}\xi}F^{\prime\prime}\big(e^{\sqrt{y}\xi}/H(y)\big)\right]}, \quad y\ge0,
\end{equation}
where $\xi\sim \N(0,1)$. Obviously, $G(0)=-F^\prime(1)/F^\pp(1)>0$ and $G$ is strictly positive and continuous on $[0,\infty)$, by Assumption \ref{ass:F:crra}.

The third function $\cG$ is defined by  
\[
\cG(x)\defeq\int_0^x \frac{1}{G(y)^2}\md y,\quad x\in[0,\infty].
\]
Obviously, $\cG(x)<\infty$ for any $x\in [0,\infty)$ and $\cG(0)=0$.

Clearly,  the three functions $H$, $G$ and $\cG$ depend only on $F$, i.e., only on  the preference functional $g$. 
{It turns out that, under some mild conditions, an equilibrium strategy is given by 
\begin{equation}\label{eq:barpi:crra:G}
\barpi_t=(\sigma^\mt (t))^{-1}\kappa(t)G\left(\cG^{-1}\left(\int_{t}^T|\kappa(s)|^2\md s\right)\right),\quad t\in[0,T)
\end{equation}
 when the market coefficients are deterministic.}

\subsection{An ansatz.}

We always assume, in the next part of Section \ref{generalCRRA}, that the market coefficients are deterministic and $k=0$ (i.e., the market is complete). In this case, $W^\Si=W$ and hence we can omit the superscripts $\Si$ and $\Oi$. 

We consider the following deterministic strategy
\begin{equation}\label{eq:pi:a}
 \barpi_t=(\sigma^\mt (t))^{-1}a_t, \quad t\in[0,T], 
 \end{equation}
where each $a_t\in\mR^d$ is deterministic and $\int_0^T|a_s|^2ds<\infty$. For this strategy, 
\begin{align}
\barX_T =\barX_t  e^{\int_t^T a_s^\mt\kappa(s)\md s -{1\over2}A(t)} R(t,T), \quad 0\le t\le T. \label{DAansatzX}
\end{align}
where $R(t,T)\defeq e^{\int_t^T a_s\cdot \md W_s}$ is the risk part of the gross rate of return ${\barX_T\over\bar X_t}$ and 
$A(t)\defeq \int_t^T |a_s|^2\md s$
is the cumulative volatility from time $t$ to $T$ of the strategy $\barpi$.

Using the positive homogeneity of $g$ and $R(t,T)$ being independent of $\F_t$, we know
\begin{equation}\label{DAansatzg}
g(\mP^t_{\barX_T})=\barX_t  e^{\int_t^T a_s^\mt\kappa(s)\md s -{1\over2}A(t)} H(A(t)),\quad t\in[0,T].
\end{equation}

\begin{proposition}\label{CRRAeqprop}
Let $g$ be given by (\ref{gCRRA}),  Assumptions \ref{ass:F:crra}--\ref{ass:P0} and \ref{ass:ui:F} be satisfied. 
Consider the deterministic strategy $\barpi$ given by \eqref{eq:pi:a}. Then, for any $p>1$, we have
\begin{equation}\label{cond:lp:xit}
\frac{\barX_T F'\left(\barX_T / g(\mP^{t}_{\barX_T})  \right)g(\mP^{t}_{\barX_T})}
{\mE_t\left[\barX_T F'\left(\barX_T / g(\mP^{t}_{\barX_T}) \right)\right]}\in L^p(\F_T),\quad \forall\,t\in[0,T),
\end{equation}
and Condition (a) of Theorem \ref{verification-pc} is equivalent to
\begin{equation}\label{CRRAFOC}
	a_t=\kappa(t)G(A(t)),\quad t\in[0,T],
\end{equation}
which implies that $A$ satisfies the following ODE
\begin{equation}\label{ode:A}
\begin{cases} A’(t)=-|\kappa(t)|^2 G(A(t))^2,\quad t\in[0,T),\\
A(T)=0.
\end{cases}
\end{equation}
\end{proposition}

\proof{Proof.} 
See Appendix \ref{sec:proof:CRRAeqprop}.
\hfill \Halmos\endproof

\vskip 5pt
In the next subsection, under some technical conditions, we are going to solve ODE \eqref{ode:A} and show that \eqref{CRRAFOC} and \eqref{eq:pi:a} give an equilibrium strategy. 

\subsection{Verification of the ansatz.}\label{verify_ansatz}

\begin{lemma}\label{lma:ode}
Let $g$ be given by (\ref{gCRRA}),  Assumptions \ref{ass:F:crra}--\ref{ass:P0} and \ref{ass:ui:F} be satisfied. 
If 
\begin{equation}\label{Gcond1}
\cG(\infty)>\int_0^T |\kappa(s)|^2 \md s,
\end{equation}
then ODE \eqref{ode:A} admits a unique solution on $[0,T]$, which is given by $A(t)=\cG^{-1}\left(\int_{t}^T|\kappa(s)|^2\md s\right)$.
\end{lemma}

\proof{Proof.}
Consider the time change $t\mapsto \bar{t}\defeq \int_{t}^T|\kappa(s)|^2\md s$, and  $\bar{A}(\bar{t})=A(t)$. Then ODE \eqref{ode:A} is equivalent to the following autonomous ODE
\begin{equation}\label{ode:barA}
\bar{A}'=G(\bar{A})^2,\quad \bar{A}(0)=0.
\end{equation}
As a consequence of Condition \eqref{Gcond1}, $\cG^{-1}$ is well defined on $\bar{t}\in \left[0,\int_0^T|\kappa(s)|^2\md s\right]$. Then $\cG^{-1}$ is the unique solution of ODE \eqref{ode:barA} based on the chain rule and the inverse function theorem. Thus, the proof follows.
\hfill \Halmos\endproof 

\vskip 5pt
Now we can conclude this subsection by the following theorem, which provides a closed-form type-I equilibrium strategy for a general CRRA betweenness preference under very mild conditions.

\begin{theorem}\label{thm:crra}
Let $g$ be given by (\ref{gCRRA}), and Assumptions \ref{ass:F:crra}--\ref{ass:P0} and \ref{ass:ui:F} as well as Condition \eqref{Gcond1} be satisfied. Then the strategy $\barpi$ given by 
\eqref{eq:barpi:crra:G} 
is a type-I equilibrium proportion strategy. 
\end{theorem}

\proof{Proof.}
See Appendix \ref{CRRAproofapp}.
\hfill \Halmos\endproof

\subsection{An example: mixed CRRA utility function.}\label{exmCRRAU}
In this subsection, we verify that all conditions in Theorem \ref{thm:crra} are satisfied if $F$ is generated by a mixture of CRRA utility functions. 

Let the CRRA betweenness preference functional $g$ be implicitly defined by 
\begin{equation}\label{mixpoweru}
	\mE\left[ \int_{-\infty}^1 U_\gamma \left(\frac{X}{g(\mP_X)}  \right)   \bF (\md \gamma)\right]=0,
\end{equation}
where each $U_\gamma$ is a CRRA utility function given by \eqref{crrau}
and $\bF$ is a probability measure on $(-\infty,1)$ such that 
$$\bF((-1/\epsilon_0,1-\epsilon_0))=1\text{ for some fixed }\epsilon_0>0.$$ 
(\ref{mixpoweru}) is a special case of (\ref{gCRRA}), with $F(x)=\int_{-\infty}^1 U_\gamma(x) \bF(\md \gamma)$, a mixture of the CRRA utility functions. 
Set $$\cP_0=\left\{\mu\in \cP((0,\infty)): \int_0^\infty x\mu(\md x)<\infty\text{ and }\int_0^\infty x^{-1/\epsilon_0}\mu(\md x)<\infty\right\}.$$

Obviously, $F$ and $\cP_0$ satisfy all of Assumptions \ref{ass:F:crra}, \ref{ass:P0} and \ref{ass:ui:F}.  It is left to verify Condition \eqref{Gcond1}. Indeed, we have 
\begin{equation}\label{mixedpowerrelation2}
		G(x)=\frac{\int_{-\infty}^1  H(x)^{-\gamma}e^{\frac{\gamma^2 x}{2}} \bF(\md \gamma)}{\int_{-\infty}^1 (1-\gamma) H(x)^{-\gamma}e^{\frac{\gamma^2 x}{2}}\bF(\md \gamma)},\quad x\in[0,\infty).	
\end{equation}
Then, ${\epsilon_0\over 1+\epsilon_0}\le G(x)\leq \frac{1}{\epsilon_0}$ for any $x\ge 0$. Therefore, \eqref{Gcond1} is now obvious. 

\vskip 5pt
\begin{remark}
(\ref{mixpoweru}) is a natural generalization of the expected CRRA utility. If $\bF$ is a Dirac measure $\delta_\gamma$, then $g(\mP_X)=U_\gamma^{-1}(\mE[U_\gamma(X)])$. In this case, \eqref{eq:barpi:crra:G} and \eqref{mixedpowerrelation2} lead to $\barpi_t=\frac{(\sigma^\mt(t))^{-1}\kappa(t)}{1-\gamma}$, which coincides  the Merton solution for the CRRA utility.
\end{remark}

\section{CARA betweenness preferences.}\label{generalCARA}

As an analogy of Section \ref{generalCRRA}, we apply Theorem \ref{verification-nc} to find equilibrium strategies for another class of betweenness preferences: the CARA betweenness preferences. We omit all proofs because they are similar to those in Section \ref{generalCRRA}.

Let $\mX=\mR$ and the preference functional $g:\cP_0\to\mR$ be implicitly defined by
\begin{equation}\label{gCARA}
	\mE [F(X-g(\mP_X))]=0,
\end{equation}
where $F$ and $\cP_0$ satisfies the following assumptions.

\begin{assumption}\label{ass:F:cara}
$F:\mR\to \mR$ is twice continuously differentiable, $F(0)=0$, $F^\prime(x)>0$ and $F^\pp(x)<0$ for all $x\in\mR$. Moreover, there exist constants $C>0$ and $\beta_0>0$ such that 
\[
|F'(x)|+|F''(x)|\leq C(1+e^{\beta_0 |x|}),\quad\forall x\in \mR.
\]
\end{assumption}
\begin{assumption}\label{ass:P0:cara}
$\cP_0$ is a nonempty convex subset of $\cP(\mR)$ satisfying the following two conditions:
\begin{description}
\item[(a)] If $\mP_X\in\cP_0$, then $\mP_{\lambda +X}\in\cP_0$ for all $\lambda\in \mR$.
\item[(b)] $\int_\mR (|F(x)|+F^\prime(x)- F^\pp(x))\mu(dx)<\infty$ for all $\mu\in\cP_0$.
 \end{description}
\end{assumption}
\begin{assumption}\label{ass:ui:F:cara}
For any $y_0\in(0,\infty)$ and $z_0\in (0,\infty)$, the family 
$$\left\{F\left(y\xi +z\right): y\in[0,y_0],\, z\in[-z_0,z_0]\right\}$$
is uniformly integrable, where $\xi\sim\N(0,1)$. 
\end{assumption}

Under Assumptions \ref{ass:F:cara}--\ref{ass:P0:cara}, using Definition \ref{derivativedef}, it is not hard to obtain
	\begin{equation*}\label{derivativeCARA}
		\nabla g(\mu,x)=\frac{F(x-g(\mu))}{\int_\mR F'(x-g(\mu))\mu(\md x)}.
	\end{equation*}
	Moreover,  Assumption \ref{gassumption1} is satisfied. In particular,  Assumption \ref{gassumption1}(c) is satisfied with the choice
	\begin{equation*}\label{CARAM1}
	M_0=0\text{ and }M_1(\mu_1,\mu_0)=\frac{\int_\mR F'(x-g(\mu_0))\mu_0(\md x)}{\int_\mR F'(x-g(\mu_0))\mu_1(\md x)}.
	\end{equation*}
Under Assumptions \ref{ass:ui:F:cara}, a function $H$ is well-defined by
\begin{equation*}\label{Hdef:cara}
\mE \left[F(  \sqrt{y}\xi -H(y)\right]=0,\quad y\geq 0,  
\end{equation*}
where $\xi\sim \N(0,1)$, and $H$ is continuous on $[0,\infty)$. We also consider the following functions:
\begin{align*}
&G(y)\defeq \frac{\mE \left[  F'(\sqrt{y}\xi-H(y)  )   \right]}{\mE[-F''(\sqrt{y}\xi-H(y))]}, \quad y\geq 0,\\
&\cG(x) \defeq \int_0^x \frac{1}{G(y)^2}\md y,\quad x \in [0,\infty].
\end{align*}
We need the following assumption on $\cG$.
\begin{assumption}\label{ass:cG:CARA}
$\cG(\infty)>\int_0^T|\kappa(s)|^2\md s$.
\end{assumption}

Now we assume that the market coefficients are deterministic. In analogy with Theorem \ref{thm:crra}, we have the following theorem.
\begin{theorem}\label{thm:CARA}
Let $g$ be given by \eqref{gCARA} and  Assumptions \ref{ass:F:cara}--\ref{ass:cG:CARA} be satisfied. 
Then the strategy $ \barpi$:
\begin{equation}\label{CARAFOC}
	 \barpi_t=(\sigma^\mt (t))^{-1}\kappa(t)G\left(\cG^{-1}\left(\int_{t}^T|\kappa(s)|^2\md s\right)\right),\quad t\in[0,T),
\end{equation}
is a type-I equilibrium dollar-amount strategy.
\end{theorem}

\begin{remark} {All conditions in Theorem \ref{thm:CARA} are satisfied when $F$ is a mixture of the CARA utility functions: $F(x)=\int_{\rho_0}^{\rho_1} (1-e^{-\rho x}) \bF(d\rho)$, where $\rho_0<\rho_1$ are two strictly positive constants and $\bF$ a probability measure on $[\rho_0,\rho_1]$.} In particular, 
if $F(x)=1-e^{-\rho x}$ for some $\rho>0$, then $g$ is the certainty equivalent of the expected CARA utility and \eqref{CARAFOC} reduces to $\barpi_t=\frac{(\sigma^\mt(t))^{-1}\kappa(t)}{\rho}$,  which coincides the Merton solution for the CARA utility.
\end{remark}

 \section{Weighted utility.}  \label{WUrandom}
 
 In this section, we mainly apply Theorem \ref{verification-pc} to investigate the equilibrium strategies for a class of weighted utilities when the market coefficients are random. We first transform the equilibrium condition \eqref{FOC} into a two-dimensional QBSDE; see \eqref{WUQBSDE} below. Then we establish the existence and uniqueness of the solution of the QBSDE; see Lemma \ref{WUQBSDE-existence}. Finally, we verify all technical conditions in Theorem \ref{verification-pc} and show that the equilibrium strategies can be obtained in terms of the solution of the QBSDE.
 
 Let  $\mX=(0,\infty)$ and a weighted utility $g:\cP_0\to(0,\infty)$ be given by 
 \begin{equation}\label{WUdef}
 g(\mP_X)=\frac{\mE \left[X^{1-\rho}X^{\gamma}\right]}{(1-\rho)\mE\left[ X^{\gamma}\right]},
 \end{equation}
 where $-1<\gamma\leq 0$, $ \gamma\le\rho< \gamma+1$ (c.f. \citet[Exercise 25.4]{Back2017}), and
 \begin{align*}
 \cP_0=\left\{\mu\in \cP((0,\infty)):\int_0^\infty (x^\gamma + x^{1-\rho+\gamma})\mu(\md x)<\infty\right\}.
 \end{align*}
  
\begin{remark} 
In \citet[pp. 661--662]{Back2017}, the preference functional of the weighted utility is defined as 
 $\left(\mE \left[X^{1-\rho}X^{\gamma}\right]/\mE\left[ X^{\gamma}\right]\right)^{1\over 1-\rho}$, which is the certainty equivalent of $X$. It is also a CRRA betweenness preference functional given by (\ref{gCRRA}) with $F(x)=x^{1-\rho}x^\gamma-x^\gamma$. Here, for notational simplicity,  we define it by \eqref{WUdef} in the \emph{utility scale}. 
\end{remark}
 
\begin{lemma}
Let $g$ be given by \eqref{WUdef}. Then Assumption \ref{gassumption1} is satisfied. Moreover, for any $x>0$ and $\mu\in \cP_0$,
 \begin{equation}\label{WU-derivative}
 	\nabla g(\mu,x)=\frac{1}{1-\rho}\cdot\frac{x^{1-\rho+\gamma}\int_0^\infty x^\gamma\mu(\md x)-x^\gamma\int_0^\infty x^{1-\rho+\gamma}\mu(\md x)}{\left(\int_0^\infty x^\gamma\mu(\md x)\right)^2}.
 \end{equation}
\end{lemma}
\proof{Proof.}
Throughout this proof we fix $\mu_0,\mu_1\in\cP_0$. For any $s\in [0,1]$ and $\mu_s=s\mu_1+(1-s)\mu_0$, we obtain from \eqref{WUdef} that
\[
g(\mu_s)=\frac{s\int_0^\infty x^{1-\rho+\gamma} \mu_1(\md x)+(1-s)\int_0^\infty x^{1-\rho+\gamma} \mu_0(\md x) }{(1-\rho)\left[s\int_0^\infty x^\gamma \mu_1(\md x)+(1-s)\int_0^\infty x^\gamma \mu_0(\md x) \right]}.
\]
Then, Conditions (a)-(b) of Assumption \ref{gassumption1} and \eqref{WU-derivative} are apparent. Moreover,
 \begin{align*}
 	g(\mu_1)-g(\mu_0)&=\frac{\left(\int_0^\infty x^{1-\rho+\gamma}\mu_1(\md x) \right)\left( \int_0^\infty x^\gamma \mu_0 (\md x)   \right)-\left(\int_0^\infty x^\gamma\mu_1(\md x) \right)\left( \int_0^\infty x^{1-\rho+\gamma} \mu_0 (\md x)   \right)}{(1-\rho)\left(\int_0^\infty x^\gamma\mu_1(\md x) \right)\left( \int_0^\infty x^\gamma \mu_0 (\md x)   \right)}\\
 	&= \frac{\int_0^\infty x^\gamma \mu_0(\md x)}{\int_0^\infty x^\gamma \mu_1(\md x)}\cdot\int_0^\infty \frac{x^{1-\rho+\gamma}\int_0^\infty x^\gamma \mu_0(\md x) - x^\gamma \int_0^\infty x^{1-\rho+\gamma}\mu_0(\md x)}{(1-\rho)\left( \int_0^\infty x^\gamma \mu_0 (\md x)   \right)^2}(\mu_1-\mu_0)(\md x)  \\
 	&=\frac{\int_0^\infty x^\gamma \mu_0(\md x)}{\int_0^\infty x^\gamma \mu_1(\md x)}\int_0^\infty \nabla g(\mu_0,x)(\mu_1-\mu_0)(\md x).
 \end{align*}
 Therefore, Assumption \ref{gassumption1}(c) is satisfied with the choice  $M_0=0$ and
 $M_1(\mu_1,\mu_0)=\frac{\int_0^\infty x^\gamma \mu_0(\md x)}{\int_0^\infty x^\gamma \mu_1(\md x)}$.\hfill \Halmos  
\endproof 
\vskip 5pt
 
When the market coefficients are deterministic, as the weighted utility is equivalent to a specific CRRA betweenness preference functional, we can find out the type-I equilibrium strategies based on the results in Section \ref{generalCRRA}. Indeed, if $F(x)=x^{1-\rho}x^\gamma-x^\gamma$, then, using \eqref{Hdef} and \eqref{Gdef}, we have, for any $y\geq 0$, $H(y)=e^{\frac{1}{2}(1-\rho+2\gamma)y}$ and $G(y)=\frac{1}{\rho-2\gamma}$. Therefore, Theorem \ref{thm:crra} implies that $\left\{ \barpi_t=\frac{(\sigma^\mt(t))^{-1}\kappa(t)}{\rho-2\gamma}, t\in [0,T] \right\} $ is a type-I equilibrium. We will revisit this result as a special case of our general results with random market coefficients; see Remark \ref{deterministicequiv}. 
 
In the next part of this section, we consider the case of random market coefficients. The randomness leads to some serious issues, which can not be solved by the methods in Sections \ref{generalCRRA}-\ref{generalCARA}. To resolve this issue, we utilize QBSDE to characterize the type-II equilibrium strategies.

For notational simplicity, we assume  $d=1$ and $k=0$.\footnote{The extension to the case $d>1$ is straightforward.}
  
 Hereafter, we denote $r_1=\gamma$, $r_2=1-\rho+\gamma$, $\lambda_1=\frac{-\gamma}{1-\rho}$, $\lambda_2=\frac{1-\rho+\gamma}{1-\rho}$. 
 
 Given a $\barpi\in H_{\rm BMO}$, define $Y_i(s)\defeq \mE_s[\barX_T^{r_i}]$, $Z_i(s)\defeq Z^{Y_i(T)}(s)$, $\hat{Y_i}(s)\defeq \log Y_i(s)$, $\hat{Z}_i(s)=Z_i(s)/Y_i(s)$, $\hat{X}_s\defeq \log \barX_s$, and $\bar{Y}^i=\hat{Y}^i-r_i\hat{X}$.   We now have the following lemma, which transforms Condition \eqref{FOC} into a QBSDE.
 
\begin{lemma}\label{lma:bsde}
 Let $g$ be given by \eqref{WUdef}. Then Condition \eqref{FOC} is equivalent to\footnote{We abuse the notation “$-i$" to represent the index other than $i$, $i\in\{1,2\}$.} 
\begin{equation}\label{WUQBSDE-1}
	\left\{
	\begin{aligned}
	\md \bar{Y}_i(s)=&-\frac{1}{2}\left[(1-r_i\lambda_i^2)\hat{Z}_i(s)^2-r_i\lambda_{-i}^2\hat{Z}_{-i}(s)^2-2r_i\lambda_{i}\lambda_{-i}\hat{Z}_{i}(s)\hat{Z}_{-i}(s)+r_i|\kappa(s)|^2\right]\md s\\
	&+\left[(1-\lambda_ir_i)\hat{Z}_i(s)-r_i\lambda_{-i}\hat{Z}_{-i}(s)-r_i\kappa(s)\right]\md W_s,\ i=1,2,\\
	\bar{Y}_1(T)=&\bar{Y}_2(T)=0.
    \end{aligned}
	\right.
\end{equation}
\end{lemma}

 \proof{Proof.} 
 See Appendix \ref{sec:lma:bsde}.
 \hfill \Halmos\endproof 
 \vskip 5pt
Using the notation $\bar{Z}_i(s)\defeq (1-\lambda_ir_i)\hat{Z}_i(s)-r_i\lambda_{-i}\hat{Z}_{-i}(s)-r_i\kappa(s)$, (\ref{WUQBSDE-1}) is further rewritten as 
\begin{equation}\label{WUQBSDE}
	\left\{
	\begin{aligned}
	\md \bar{Y}_i(s)=&-\frac{1}{2}\left[\bar{Z}(s)^\mt \bC^i\bar{Z}(s)+\bc_{i,i}\bar{Z}_i(s)\kappa(s)+\bc_{-i,i}\bar{Z}_{-i}\kappa(s)+\bfb_i|\kappa(s)|^2\right]\md s\\
		&+\bar{Z}_i(s) \md W_s,\ i=1,2,\\
	\bar{Y}_1(T)=&\bar{Y}_2(T)=0,
	\end{aligned}
    \right.
\end{equation}
here, $\bar{Z}=\left(\begin{array}{c}\bar{Z}_1\\ \bar{Z}_2\end{array}\right)$, and $\bC^1, \bC^2\in \mR^{2\times 2}$, $\bc_{i,j},\bfb_i\in \mR$ for $i,j=1,2$. The calculations of $\bC^i$, $\bc_{i,j}$, and $\bfb_i$ are tedious but irrelevant to the main results. Therefore we put them in Appendix \ref{WUQBSDEcoefficient}.

\vskip 5pt

The next task is to establish the solvability of (\ref{WUQBSDE}). To this end, let $\Theta\defeq \int_0^T|\kappa(s)|^2\md s$ and $V(\Theta)\defeq \sup_\tau\|\Theta-\mE_\tau [\Theta]   \|_{\infty}$.

\begin{lemma}\label{WUQBSDE-existence}
	For any sufficiently small $\varrho>0$, there exists $V_0>0$ such that if $V(\Theta)<V_0$, then the QBSDE (\ref{WUQBSDE}) admits a unique solution $(\bar{Y},\bar{Z})\in (L^{\infty}(\mF,\mR))^2\times (H^d_{\rm BMO})^2$ with $\|\bar{Z}\|_{\rm BMO}<\varrho$. 
\end{lemma}
\proof{Proof.}
See Appendix \ref{proofQBSDEexistence}.
\hfill \Halmos\endproof

\vskip 5pt
\begin{remark} 
\begin{itemize}
    \item[(a)] It is worth noting that (\ref{WUQBSDE}) is a \emph{two-dimensional} QBSDE, whose two components $\bar{Y}_1$ and $\bar{Y}_2$ are coupled. Two-dimensional QBSDEs are technically much more challenging than the one-dimensional ones, and unless other assumptions are imposed, the well-posedness remains open; see the bibliographical notes in Subsection 7.4 of \cite{Zhang2017}. Furthermore, existing results on multi-dimensional QBSDE with diagonally quadratic generators (\cite{Hu2016} and \cite{Fan2023}) or triangular quadratic generators (\cite{Jackson2022}) are not applicable to \eqref{WUQBSDE}. 
\item[(b)]
   To obtain the well-posedness, or at least the existence of solution, we need certain \emph{smallness} conditions on the terminal value or other parameters, which is fairly common in literature; see, e.g., \cite{Tevzadze2008}, \cite{Frei2014}, \cite{Kramkov2016}, \cite{Jamneshan2017}, \cite{Herdegen2021}, \cite{Fu2023}, among others. In our example, we need to impose such assumptions on $\kappa$. One possible resolution is that we assume that $\|\kappa\|_\infty$ is sufficiently small. However, recalling that $\kappa$ is the market price of risk, this assumption is by no means reasonable economically. We resolve this issue by assuming that $V(\Theta)$ is small. In other words, we assume that $\kappa$ is close to a deterministic constant. Clearly, when $\Theta$ is deterministic, $V(\Theta)=0$, and vice versa. 
\end{itemize}
\end{remark}

Finally, with Lemma \ref{WUQBSDE-existence}, we establish the verification result for this example. 

\begin{proposition}\label{verificationWU-random}
	Let $g$ be given by \eqref{WUdef} and  $(\bar{Y},\bar{Z})$ be given by Lemma \ref{WUQBSDE-existence}. Assume  that $V(\Theta)$ is sufficiently small. Then a Type-II equilibrium is given by
	\begin{equation}\label{WUrelationtransform}
	\sigma(t)\barpi_t=\frac{1}{\rho-2\gamma}\kappa(t)+\frac{1}{\rho-2\gamma}[\lambda_1\bar{Z}_1(t)+\lambda_2\bar{Z}_2(t)]. 
\end{equation}
\end{proposition}
\proof{Proof.}
	See Appendix \ref{proofverificationWU}.
\hfill \Halmos\endproof

\vskip 5pt
\begin{remark}\label{deterministicequiv}
	If $\kappa$ is deterministic, (\ref{WUQBSDE}) implies $\bar{Z}_1=\bar{Z}_2=0$. We then observe from (\ref{WUrelationtransform}) that $\barpi_t=\frac{\kappa(t)}{\sigma(t)(\rho-2\gamma)}$. In other words, in the deterministic market coefficient case, the weighted utility agent, with weighting exponent $\gamma$ and risk aversion $\rho$, is equivalent to an expected utility agent, with risk aversion $\rho-2\gamma$. Therefore, our result provides an additional explanation of the claim in \cite{Back2017}, asserting that “the relative risk aversion of CRRA weighted utility should be regarded as $\rho-2\gamma$" (see the last line of page 661 therein).
\end{remark}

\section{Conclusion.}\label{conclu}
{
In conclusion, this paper makes contributions to the study of dynamic portfolio selection problems with nonlinear law-dependent preferences. We propose a notion of derivatives with respect to distribution, and prove verification theorems based on the method of stochastic maximum principle. We utilize our verification theorems to investigate several concrete cases. 
Specifically, in the cases of CRRA and CARA betweenness preferences with deterministic market coefficients, we provide the equilibrium strategies in closed form. In the case of weighted utility with random market coefficients, the equilibrium strategy is described by a coupled system of QBSDEs.
}
\vskip 5pt

\section*{Acknowledgments.}
This research is supported by the National Key R\&D Program of China (NO. 2020YFA0712700) and NSFC (NOs.
12071146, 12271290, 11871036).

\begin{APPENDIX}{}

\section{Proofs of the verification theorems.}\label{vthmproof}

\subsection{Proof of Theorem \ref{verification-pc}.}\label{proof:verification-pc}
To prepare for the proof of Theorem \ref{verification-pc}, we need two lemmas.

\begin{lemma}\label{adjoint}
Let $p>1$. Fix $t\in [0,T)$ and $\barpi\in H^d_{\rm BMO}$. Suppose that $\xi^t$ is an $\F_T$-measurable random variable such that $\barX_T\xi^t \in L^p(\F_T)$. Define
	\begin{align*}
		p^t(s)\defeq & \barX_s^{-1}\mE_s[\barX_T \xi^t],\\
		q^{t,\Si}(s)\defeq & \barX_s^{-1}Z^{\barX_T\xi^t,\Si}(s)-\sigma(s)^\mt\barpi_s p^t(s),\\
		q^{t,\Oi}(s)\defeq & \barX_s^{-1}Z^{\barX_T\xi^t,\Oi}(s).
	\end{align*}
	Then $(p^t,q^{t,\Si},q^{t,\Oi})$ satisfies the following BSDE
	\begin{equation}\label{adjointeq}
		\left\{
		\begin{aligned}
			&\md p^t(s)=-\barpi_s^\mt(\theta(s)p^t(s)+\sigma(s)q^{t,\Si}(s))\md s+q^{t,\Si}(s)\cdot\md W^{\Si}_s+q^{t,\Oi}(s)\cdot\md W^{\Oi}_s,\\
			&p^t(T)=\xi^t.	
		\end{aligned}
		\right.	
	\end{equation}
Moreover, $\barX p^t\in S^p(\mF,\mR)$, $\barX q^{t,\Si}\in L^{p_1,2}(\mF,\mR^d)$ for every $p_1\in (1,p)$, and $\barX q^{t,\Oi}\in L^{p,2}(\mF,\mR^k)$. 
\end{lemma}

\proof{Proof.}
	 The equation (\ref{adjointeq}) is directly derived from It$\hato$'s formula. Because $\barX_T \xi^t\in L^p(\F_T)$, based on Lemma \ref{MRM-p}, we know $\barX p^t\in S^p(\mF,\mR)$, $\barX q^{t,\Oi}\in L^{p,2}(\mF,\mR^k)$. To show $\barX q^{t,\Si}\in L^{p_1,2}(\mF,\mR^d)$, we only need to prove $\sigma^\mt \barpi \barX p^t\in L^{p_1,2}(\mF,\mR^d)$. Actually, 
	\begin{align*}
		\mE\left[\left(\int_0^T |\sigma(s)^\mt\barpi_s|^2|\barX_sp^t(s)|^2\md s  \right)^{p_1/2} \right]&\leq \|\sigma\|_{\infty}\mE\left[\sup_{s\in [0,T]}|\barX_s p^t(s)|^{p_1}\left(\int_0^T|\barpi_s|^2\md s  \right)^{p_1/2}   \right]\\
		&\leq \|\sigma\|_{\infty}\|\barX p^t\|^{p_1}_{S^p(\mF,\mR)}\left\{\mE\left[\left( \int_0^T |\barpi_s|^2\md s    \right)^{\frac{p_1p}{2(p-p_1)}} \right]        \right\}^{1-p_1/p}\\
		&<\infty.
	\end{align*}
	Here, we have used the fact that $\mE\left[\left(\int_0^T|\barpi_s|^2\md s\right)^n\right]<\infty$ for all $n\geq 1$,
	which is a consequence of $\barpi\in H^d_{\rm BMO}$; see  \citet[(7.2.6)]{Zhang2017}. 
\hfill \Halmos\endproof 

For the next lemma we use the following notation: for $\varphi\in L^{\infty}(\F_t,\mR^d)$,
\begin{equation}\label{Epsilondef}
\E^\varphi_s\defeq \exp\left(\int_0^s (\varphi^\mt \theta(r)-\varphi^\mt \sigma(r)\sigma(r)^\mt \barpi_r-\frac{1}{2}|\varphi^\mt \sigma(r)|^2)\md r +\int_0^s \varphi^\mt \sigma(r)\md W^{\Si}_r  \right).
\end{equation}
We now consider the integrability of $\E^\varphi$.
\begin{lemma}\label{Eintegrallemma}
	If $\barpi\in H^d_{\rm BMO}$, then, for any $p>1$, we have
	\begin{equation}\label{Eintegraleq}
		\mE \left[ \sup_{s\in [0,T]}(|\E^\varphi_s|^p+|\E^\varphi_s|^{-p})\right]<\infty.	
	\end{equation}
\end{lemma}
\proof{Proof.}
	We only prove $\mE\left[ \sup_{s\in [0,T]}|\E^\varphi_s|^p\right]<\infty$, as the rest part can be similarly proved. 
\begin{align*}
(\E^\varphi_s)^p
 =\exp\left\{\int_0^s p(\varphi^\mt \theta(r)-\varphi\sigma(r)\sigma(r)^\mt \barpi_r-\frac{1}{2}|\varphi^\mt \sigma(r)|^2)\md r   
 +\frac{1}{2}\int_0^sp^2|\varphi^\mt\sigma(r)|^2 \md r  \right\}\cdot \cZ^{W^{\Si}}_s(p\varphi^\mt \sigma).
 \end{align*}
Because $\varphi,\theta,\sigma$ are all bounded, we have, for some constant $C>0$ and any sufficiently small $c>0$,
\begin{align*}
\mE^{\mQ}\left[\sup_{0\leq s\leq T}(\E^\varphi_s)^p\right]&\leq C \left( \mE \left[e^{2p\|\varphi\|_{\infty}\|\sigma\|_{\infty}^2 \int_0^T |\barpi_r|\md r} \right] \right)^{1/2}\cdot \left(\mE \left[\sup_{0\leq s\leq T}(\cZ^{W^{\Si}}_s(p\varphi^\mt \sigma))^2    \right]  \right)^{1/2}\\
&\leq C\left( \mE \left[e^{4p^2\|\varphi\|_{\infty}^2\|\sigma\|_{\infty}^4T/c+c\int_0^T|\barpi_r|^2\md r}  \right]    \right)^{1/2}\\
&<\infty.
\end{align*}
Here, we have used the energy inequality of BMO martingales (see, \citet[Lemma C.1]{Fu2023}) to conclude  $\mE\left[  e^{c\int_0^T|\barpi_s|^2\md s}     \right]<\infty$ for any sufficiently small $c>0$.
\hfill \Halmos\endproof 
\vskip 5pt
\proof{Proof of Theorem \ref{verification-pc}.}
Let $t\in [0,T)$ and $\varphi\in L^{\infty}(\F_t,\mR^d)$ be fixed. 

For sufficiently small $\epsilon>0$, by Assumption \ref{gassumption1} and Condition (a), we have 
	\begin{align*}
		g(\mP^{t}_{\barX^{t,\epsilon,\varphi}_T})-g(\mP^{t}_{\barX_T})&\leq M_1(\mP^{t}_{\barX^{t,\epsilon,\varphi}_T},\mP^{t}_{\barX_T})\int_\mR \nabla g(\mP^{t}_{\barX_T},x)(\mP^{t}_{\barX^{t,\epsilon,\varphi}_T}-\mP^{t}_{\barX_T})(\md x) +M_0(\mP^{t}_{\barX^{t,\epsilon,\varphi}_T},\mP^{t}_{\barX_T}) \\
		&=M_1(\mP^{t}_{\barX^{t,\epsilon,\varphi}_T},\mP^{t}_{\barX_T})\mE_t\left[\nabla g(\mP^{t}_{\barX_T},\barX^{t,\epsilon,\varphi}) -\nabla g(\mP^{t}_{\barX_T},\barX_T)\right]+M_0(\mP^{t}_{\barX^{t,\epsilon,\varphi}_T},\mP^{t}_{\barX_T})\\
		&\leq M_1(\mP^{t}_{\barX^{t,\epsilon,\varphi}_T},\mP^{t}_{\barX_T})\mE_t\left[\xi^t(\barX^{t,\epsilon,\varphi}_T-\barX_T)\right]+M_0(\mP^{t}_{\barX^{t,\epsilon,\varphi}_T},\mP^{t}_{\barX_T}).
	\end{align*}
	Based on Lemma \ref{adjoint} (in particular (\ref{adjointeq})) and It$\hato$'s formula, we have
	\begin{align*}
		\xi^t (\barX^{t,\epsilon,\varphi}_T-\barX_T)=&-\int_t^T (\barX^{t,\epsilon,\varphi}_s-\barX_s)\left(\barpi_s^\mt (\theta(s)p^t(s)+\sigma(s)q^{t,\Si}(s))\md s
		+q^{t,\Si}\cdot \md W^{\Si}_s+q^{t,\Oi}\cdot \md W^{\Oi}_s\right)\\
		&+\int_t^T p^t(s)\left(\barX^{t,\epsilon,\varphi}_s(\barpi^{t,\epsilon,\varphi}_s)^\mt -\barX_s\barpi_s\right)(\theta(s)\md s+\sigma(s)\md W^{\Si}_s)\\
		&-\int_t^T\left(\barX^{t,\epsilon,\varphi}_s(\barpi^{t,\epsilon,\varphi}_s)^\mt -\barX_s\barpi_s^\mt\right)\sigma(s)q^{t,\Si}(s)\md s\\
		=&\int_t^T\left((\barX^{t,\epsilon,\varphi}_s-\barX_s)q^{t,\Si}(s) + p^t(s)\sigma(s)^\mt(\barX^{t,\epsilon,\varphi}_s\barpi^{t,\epsilon,\varphi}_s -\barX_s\barpi_s)\right)\cdot\md W^{\Si}_s \\
		&+\int_t^T (\barX^{t,\epsilon,\varphi}_s-\barX_s)q^{t,\Oi}(s)\cdot \md W^{\Oi}_s\\
		&-\varphi^\mt \int_t^{t+\epsilon}(\barX^{t,\epsilon,\varphi}_s-\barX_s)(p^t(s)\theta(s)+\sigma(s)q^{t,\Si}(s)) \md s\\
		&-\varphi^\mt \int_t^{t+\epsilon}\barX_s(p^t(s)\theta(s)+\sigma(s)q^{t,\Si}(s)) \md s\\
		\eqdef& \int_t^T \Ui^{t,\Si}(s)\cdot \md W^{\Si}(s)+\int_t^T\Ui^{t,\Oi}(s)\cdot \md W^{\Oi}(s)+\varphi^\mt \Gamma^{t,\epsilon,\varphi}+\varphi^\mt \Lambda_\epsilon(t),
	\end{align*}
	where
	\begin{align*}
		&\Gamma^{t,\epsilon,\varphi}=-\int_t^{t+\epsilon}(\barX^{t,\epsilon,\varphi}_s-\barX_s)(p^t(s)\theta(s)+\sigma(s)q^{t,\Si}(s)) \md s,\\
		&\Lambda_\epsilon(t)=-\int_t^{t+\epsilon}\barX_s(p^t(s)\theta(s)+\sigma(s)q^{t,\Si}(s)) \md s.
	\end{align*}
	It is clear that $\barX^{t,\epsilon,\varphi}_s=\barX_s \E^\varphi_{s\wedge (t+\epsilon)}/\E^\varphi_t$ for $s\geq t$, where (see (\ref{Epsilondef}))
	\[
	\E^\varphi_s =\exp\left(\int_0^s (\varphi^\mt \theta(r)-\varphi^\mt \sigma(r)\sigma(r)^\mt \barpi_r-\frac{1}{2}|\varphi^\mt \sigma(r)|^2)\md r +\int_0^s \varphi^\mt \sigma(r)\md W^{\Si}_r  \right).
	\]
	Therefore, for $s\geq t$, we have 
	\begin{align*}
		\Ui^{t,\Si}(s)&=p^t(s)\sigma(s)(\barX^{t,\epsilon,\varphi}\varphi\ind_{[t,t+\epsilon)})+(\barX^{t,\epsilon,\varphi}_s-\barX_s)(p^t(s)\sigma(s)^\mt \barpi+q^{t,\Si}(s)) \\
		&=p^t(s)\barX_s \E^\varphi_{t+\epsilon}\varphi\ind_{[t,t+\epsilon)}/\E^\varphi_t+(\E^\varphi_{s\wedge(t+\epsilon)}/\E^\varphi_t-1)Z^{t,\Si}(s).
	\end{align*}
	Because $\barX p^t\in S^p(\mF,\mR)$, $Z^{t,\Si}\in L^{p,2}(\mF,\mR^d)$ and $\mE_t\left[\sup_{s\in [0,T]}\left((\E^\varphi_s)^n+(\E^\varphi_s)^{-n}\right)\right]<\infty$, for any $n\geq 1$, based on Lemma \ref{Eintegrallemma}, it is not hard to show $\Ui^{t,\Si}\in L^{p',2}(\mF,\mR^d)$ for some $p’\in(1,p)$. Using the BDG inequality, we know that $\int_0^\cdot \Ui^{t,\Si}(s)\cdot\md W^{\Si}_s$ is a martingale. Similarly,  $\int_0^\cdot \Ui^{t,\Oi}(s)\cdot\md W^{\Oi}_s$ is also a martingale. Hence
	\[
	\mE_t\left[\int_t^T \left(\Ui^{t,\Si}(s)\cdot \md W^{\Si}_s+\Ui^{t,\Oi}(s)\cdot \md W^{\Oi}_s \right)\right]=0.
	\]
	
	From the above discussion, we have
	\begin{equation}\label{gperturb}
		g(\mP^{t}_{\barX^{t,\epsilon,\varphi}_T})-g(\mP^{t}_{\barX_T})\leq M_1(\mP^{t}_{\barX^{t,\epsilon,\varphi}_T},\mP^{t}_{\barX_T})\left( \varphi^\mt\mE_t[\Gamma^{t,\epsilon,\varphi}]+\varphi^\mt \mE_t[\Lambda_\epsilon(t)] \right)+M_0(\mP^{t}_{\barX^{t,\epsilon,\varphi}_T},\mP^{t}_{\barX_T}).
	\end{equation}
	
	We now deal with $\Gamma^{t,\epsilon,\varphi}$. To this end, we first estimate $\E^\varphi_{t+\epsilon}-\E^\varphi_t$. By It$\hato$'s formula, 
	\[
	\E^\varphi_{t+\epsilon}-\E^\varphi_t=\int_t^{t+\epsilon}\E^\varphi_s(\varphi^\mt \theta(s)-\varphi^\mt\sigma(s)\sigma(s)^\mt \barpi_s)\md s +\int_t^{t+\epsilon}\E^\varphi_s \varphi^\mt \sigma(s)\md W^{\Si}_s.
	\]
	Therefore, for any $m\geq 1$, we have
	\begin{equation}\label{Eincrementest}
	\begin{aligned}
		\{\mE_t |\E^\varphi_{t+\epsilon}-\E^\varphi_t|^m\}^{1/m}\leq& C\left\{\int_t^{t+\epsilon}(\mE_t[|\E^\varphi_s|^m])^{1/m} \md s+ \left(\mE_t \left[\left(\int_t^{t+\epsilon}|\E^\varphi_s||\barpi_s|\md s    \right)^m \right]    \right)^{1/m} \right.\\
		&\left. +\left(\mE_t\left[\left(\int_t^{t+\epsilon}\E^\varphi_s\varphi^\mt \sigma(s)\md W^{\Si}_s\right)^m\right] \right)^{1/m}                \right\}\\
		\leq &C_m\left\{\epsilon+\sqrt{\epsilon}\left(\mE_t\left[\sup_s|\E^\varphi_s|^m\left(\int_0^T |\barpi_s|^2\md s  \right)^{m/2} \right]  \right)^{1/m}\right. \\
		&\left.+\left(\mE_t\left[\left(\int_t^{t+\epsilon} |\E^\varphi_s|^2\md s\right)^{m/2}  \right] \right)^{1/m}\right\}\\
		\leq &C_m \sqrt{\epsilon}.
	\end{aligned}
\end{equation}
	Here, $C_m$ (or $C$ in the rest of the proof) is a almost surely finite random variable, which does not depend on $\epsilon$. Then
	\begin{align*}
	\mE_t[|\Gamma^{t,\epsilon,\varphi}|]\leq& C\mE_t\left[\int_t^{t+\epsilon}|\barX^{t,\epsilon,\varphi}-\barX_s|(|p^t(s)+|q^{t,\Si}(s)|) \md s \right]  \\
		=&C \mE_t\left[|\E^\varphi_{t+\epsilon}-\E^\varphi_t||\E^\varphi_t|^{-1} \left(\int_t^{t+\epsilon} \left(|\barX_sp^t(s)|+|\barX_sq^t(s)|\right)\md s \right)\right]\\
		\leq& C\epsilon \mE_t\left[|\E^\varphi_{t+\epsilon}-\E^\varphi_t||\E^\varphi_t|^{-1}\sup_s|\barX_sp^t(s)|   \right] \\
		 &+C\sqrt{\epsilon} \left\{\mE_t\left[|\E^\varphi_{t+\epsilon}-\E^\varphi_t||\E^\varphi_t|^{-1}\left(\int_t^{t+\epsilon}|\barX_s q^t(s)|^2\md s \right)^{1/2}\right]\right\}.
	\end{align*}
	 Appropriately choosing $p_1,p_2,p_3,p_4>1$ and $1<p'<p$ (recall the $L^p$-integrability of $\barX_T\xi^t$) and using H$\mathrm{\ddot{o}}$lder's inequality, we obtain
	\begin{align*}
		\mE_t[|\Gamma^{t,\epsilon,\varphi}|]\leq &C\left( \epsilon \left(\mE_t[|\E^\varphi_{t+\epsilon}-\E^\varphi_t|^{p_1}]\right)^{1/p_1}\left(\mE_t[(\E^\varphi_t)^{-p_2} ]\right)^{1/p_2}\left(\mE_t[\sup_s|\barX_sp^t(s)|^p]\right)^{1/p}\right.\\
		&\left.+\sqrt{\epsilon}\left(\mE_t[|\E^\varphi_{t+\epsilon}-\E^\varphi_t|^{p_3}]\right)^{1/p_3}\left(\mE_t[(\E^\varphi_t)^{-p_4} ]\right)^{1/p_4}\left(\mE_t\left[\left(\int_t^{t+\epsilon}|\barX_sq^t(s)|^2\md s\right)^{p'/2}\right]\right)^{1/p'}   \right)\\
		=& o(\epsilon).
	\end{align*}

	To proceed, we observe that \eqref{FOC} is equivalent to 
\begin{equation}\label{FOCproof}
\theta(s)p^s(s)+\sigma(s)q^{s,\Si}(s)=0,\quad\forall t\in [0,T].
\end{equation}

	If we have (a), (b) and (c), we conclude from (\ref{Mbound}), (\ref{Nbound}) and (\ref{gperturb}) that, for {\it any} $t\in [0,T)$ and $\varphi\in L^{\infty}(\F_t,\mR^d)$,
	\begin{equation}\label{incrementsest}
	\limsup_{\epsilon\to 0 }\left[\frac{g\left(\mP^{t}_{\barX^{t,\epsilon,\varphi}_T}\right)-g\left(\mP^{t}_{\barX_T}\right)}{\epsilon} \right]\leq \limsup_{\epsilon\to 0}\frac{1}{\epsilon} M_1(\mP^{t}_{\barX^{t,\epsilon,\varphi}_T},\mP^{t}_{\barX_T})\varphi^\mt \mE_t[\Lambda_\epsilon(t)].
	\end{equation}
	By using Condition (b), (\ref{Mbound}), (\ref{FOCproof}), and the Dominated Convergence Theorem, it is not hard to show 
	\[
	\lim_{\epsilon\to }\frac{1}{\epsilon}\mE_t |M(\mP^{t}_{\barX^{t,\epsilon,\varphi}_T},\mP^{t}_{\barX_T})\varphi^\mt \Lambda_\epsilon(t)|=0.
	\]
	 Therefore $\barpi$ is a Type-I equilibrium. 
	\vskip 5pt
	We now consider the case where (b') and (c'), instead of (b) and (c), hold. Repeating similar estimates as in (\ref{Eincrementest}) and using Lemma \ref{Eintegrallemma}, we obtain, for any $m\geq 1$,
	\[
	\left(\mE[|\E^\varphi_{t+\epsilon}-\E^\varphi_t|^m]\right)^{1/m}\leq K_m\sqrt{\epsilon}.
	\]
	Here $K_m$ and $K$ in the rest of this proof are positive constants that may vary line to line. Therefore, appropriately choosing $p_5,p_6,p_7,p_8>1$ and $1<p''<p'<p$, we conclude
	\begin{align*}
		\left(\mE[|\Gamma^{t,\epsilon,\varphi}|^{p''}]\right)^{1/p''}\leq &K\left( \epsilon \left(\mE[|\E^\varphi_{t+\epsilon}-\E^\varphi_t|^{p_5}]\right)^{1/p_5}\left(\mE[(\E^\varphi_t)^{-p_6} ]\right)^{1/p_6}\left(\mE\left[\sup_s|\barX_sp^t(s)|^p\right]\right)^{1/p}\right.\\
		&\left.+\sqrt{\epsilon}\left(\mE[|\E^\varphi_{t+\epsilon}-\E^\varphi_t|^{p_7}]\right)^{1/p_7}\left(\mE[(\E^\varphi_t)^{-p_8}] \right)^{1/p_8}\left(\mE\left[\left(\int_t^{t+\epsilon}|\barX_sq^t(s)|^2\md s\right)^{p'/2}\right]\right)^{1/p'}   \right)\\
		=& o(\epsilon).
	\end{align*}
	Invoking (\ref{Nintegral}), we derive from (\ref{gperturb}) that, for any $t\in [0,T)$ and any nonnegative $\zeta\in L^{\infty}(\F_t)$,
	\begin{equation}\label{incrementsest2}
		\limsup_{\epsilon\to 0 }\mE\left[\frac{g\left(\mP^{t}_{\barX^{t,\epsilon,\varphi}_T}\right)-g\left(\mP^{t}_{\barX_T}\right)}{\epsilon}\cdot \zeta \right]\leq \limsup_{\epsilon\to 0}\frac{1}{\epsilon} \mE\left[ M_1(\mP^{t}_{\barX^{t,\epsilon,\varphi}_T},\mP^{t}_{\barX_T})\zeta \varphi^\mt\Lambda_\epsilon(t)\right].
	\end{equation}
	Noting that 
	\[
	\barX_s(\theta(s) p^t(s)+\sigma(s)q^{t,\Si}(s))=D(t)\sum_{i=1}^M A_i(s)B_i(s)C_i(t),\] 
	(\ref{FOCproof}) is equivalent to 
	\[
	\sum_{i=1}^{M}A_i(s)B_i(s)C_i(s)=0. 
	\]
	Thus, for sufficiently small $m_0>0$, $m\defeq 1+m_0$ and $l,n>1$ being as appropriate, we have
	\begin{align*}
		&\left|\mE \left[ M_1(\mP^{t}_{\barX^{t,\epsilon,\varphi}_T},\mP^{t}_{\barX_T})\zeta \varphi^\mt\Lambda_\epsilon(t)\right]\right| \\
		\leq& \mE\left[ \vphantom{\sum_{k=0}^M\int_0^1} \left|M_1(\mP^{t}_{\barX^{t,\epsilon,\varphi}_T},\mP^{t}_{\barX_T})\right|\|\zeta \varphi\||D(t)| \times \sum_{i=1}^M\int_t^{t+\epsilon} |A_i(s)||B_i(s)| |C_i(t)-C_i(s)|\md s\right]\\
		\leq& K\mE \left[ \vphantom{\sum_{k=0}^M\int_0^1}   \left|M_1(\mP^{t}_{\barX^{t,\epsilon,\varphi}_T},\mP^{t}_{\barX_T})\right| \times \sum_{i=1}^M\sup_{t\leq s\leq t+\epsilon}|D(t)||B_i(s)||C_i(t)-C_i(s)| \cdot  \int_t^{t+\epsilon}|A_i(s)|\md s\right]\\
		\leq& \sup_{i=1,\cdots,M}MK\left(\mE\left[|M_1(\mP^{t}_{\barX^{t,\epsilon,\varphi}_T},\mP^{t}_{\barX_T})|^n  \right]  \right)^{1/n} \times \left(\mE \left[\left( \int_t^{t+\epsilon}|A_i(s)|\md s  \right)^l   \right]\right)^{1/l}\\
		& \times\left(\mE \left[\sup_{t\leq s\leq t+\epsilon}(|D(t)||B_i(s)||C_i(t)-C_i(s)|)^m\right]  \right)^{1/m}	\\
		\leq& K\epsilon \left(\mE\left[ \left( \int_t^{t+\epsilon}|A_i(s)|^2\md s  \right)^{l/2}  \right] \right)^{1/l}	\\
		=& o(\epsilon).
	\end{align*}
	This, combined with (\ref{incrementsest2}), yields that $\barpi$ is a Type-II equilibrium.
\hfill \Halmos\endproof

\subsection{Proof of Theorem \ref{verification-nc}.}\label{proof:verification-nc}
	Throughout this proof, we fix $t\in [0,T)$ and $\varphi\in L^{\infty}(\F_t,\mR^d)$. Slightly abusing the notation, we denote $p^t=\{\mE_s[\xi^t],s\in [0,T] \}$ and $q^t=Z^{\xi^t}$ in this proof.
	
	By Assumption \ref{gassumption1},
	\[
	g(\mP^{t}_{\barX^{t,\epsilon,\varphi}_T})-g(\mP^{t}_{\barX_T})\leq M_1(\mP^{t}_{\barX^{t,\epsilon,\varphi}_T},\mP^{t}_{\barX_T})\mE_t\xi^t(\barX^{t,\epsilon,\varphi}_T-\barX_T)+M_0(\mP^{t}_{\barX^{t,\epsilon,\varphi}_T},\mP^{t}_{\barX_T}).
	\]
	We note from (\ref{wealthdynamic:amount}) that
	\begin{align*}
	& \barX_s=\barX_0+\int_0^s \barpi_s^\mt(\theta(r)\md r +\sigma(r)\cdot\md W^{\Si}_r), \\
	& \barX^{t,\epsilon,\varphi}_s-\barX_s=\int_0^s\ind_{[t,t+\epsilon)}(r)\varphi^\mt (\theta(r)\md r+\sigma(r)\cdot\md W^{\Si}_r).	
	\end{align*}
 Let	$p^t\defeq \{\mE_s[\xi^t],s\in [0,T]\}$ and $q^t\defeq Z^{\xi^t}$.
It follows from It$\hato$'s formula that
	\begin{align*}
	\xi^t(\barX^{t,\epsilon,\varphi}_T-\barX_T)
	=&\int_t^T (\barX^{t,\epsilon,\varphi}_s-\barX_s)(q^{t,\Si}(s)\cdot \md W^{\Si}_s+q^{t,\Oi}(s)\cdot \md W^{\Oi}_s)\\
	&+\varphi^\mt\int_t^{t+\epsilon}p^t(s)(\theta(s)\md s+\sigma(s)\cdot\md W^{\Si}_s)+\varphi^\mt\int_t^{t+\epsilon}\sigma(s)q^{t,\Si}(s)\md s.	
	\end{align*}
It is then not hard to show $\mE_t\xi^t[(\barX^{t,\epsilon,\varphi}_T-\barX_T)]=\varphi^\mt\mE_t\left[\int_t^{t+\epsilon}(\theta(s)p^t(s)+\sigma(s)q^t(s))\md s\right]$. Using (\ref{Mbound}), (\ref{Nbound}) and Conditions (a) and (b), we obtain
\begin{align*}
	\limsup_{\epsilon\to 0}\frac{1}{\epsilon} [g(\mP^{t}_{\barX^{t,\epsilon,\varphi}_T})-g(\mP^{t}_{\barX_T})]&\leq \limsup_{\epsilon\to 0} |M_1(\mP^{t}_{\barX^{t,\epsilon,\varphi}_T},\mP^{t}_{\barX_T})| \cdot
\mE_t\left[\left|   \frac{\int_t^{t+\epsilon}(\theta(s)p^t(s)+\sigma(s)q^t(s))\md s}{\epsilon}\right|\right]\\
	&\leq \limsup_{\epsilon\to 0}  |M_1(\mP^{t}_{\barX^{t,\epsilon,\varphi}_T},\mP^{t}_{\barX_T})| \cdot  \mE_t[|\theta(t)p^t(t)+\sigma(t)q^t(t)| ]   \\
	&= 0.
\end{align*}
Thus $\barpi$ is a Type-I equilibrium.\hfill \Halmos

\section{Proofs for Section \ref{generalCRRA}.}

\subsection{Proof of Proposition \ref{CRRAeqprop}.}\label{sec:proof:CRRAeqprop}
First, using Assumption \ref{ass:F:crra}, \eqref{DAansatzX} and \eqref{DAansatzg}, we derive
\begin{align*}
\frac{\barX_T F'\left(\barX_T / g(\mP^{t}_{\barX_T})  \right)g(\mP^{t}_{\barX_T})}
{\mE_t\left[\barX_T F'\left(\barX_T / g(\mP^{t}_{\barX_T}) \right)\right]}&=\frac{\barX_T F'\left(R(t,T)/H(A(t))  \right)}
{\mE_t\left[\frac{R(t,T)F'\left(R(t,T)/H(A(t)) \right)}{H(A(t))}\right]}\\
&=\frac{\barX_T H(A(t)) F'(R(t,T)/H(A(t)))}{\mE\left[e^{\sqrt{A(t)}\xi}F'(e^{\sqrt{A(t)}\xi}/H(A(t))     \right]}\\
&\leq \frac{CH(A(t))\barX_T}{\mE\left[e^{\sqrt{A(t)}\xi}F'(e^{\sqrt{A(t)}\xi}/H(A(t))     \right]}\cdot \left(  \left(\frac{R(t,T)}{H(A(t))}\right)^{\alpha_0} +\left(\frac{R(t,T)}{H(A(t))}\right) ^{-\alpha_0} \right)\\
&\in L^p,\quad\forall p>1,
\end{align*}
because $A(t)$ is deterministic and both of $R(t,T)$ and $\barX_T$ have log-normal distributions. Thus, we have \eqref{cond:lp:xit}.

Based on  Lemma \ref{CRRAderivativelemma}, $\nabla g(\mu,\cdot)$ is smooth. In view of Condition (a) of Theorem \ref{verification-pc}, we have
\[
\barX_T\xi^t=\barX_T\partial_x \nabla g(\mP^{t}_{\barX_T},\barX_T)
=\frac{\barX_T F'\left(\barX_T / g(\mP^{t}_{\barX_T})  \right)g(\mP^{t}_{\barX_T})}
{\mE_t\left[\barX_T F'\left(\barX_T / g(\mP^{t}_{\barX_T}) \right)\right]}.
\] 
Using Condition (\ref{FOC}) yields
\begin{equation}\label{CRRAFOC_ZE}
a_t=\kappa(t)+\frac{Z^{\barX_TF'\left(\barX_T / g(\mP^{t}_{\barX_T}) \right)}(t)}
{\mE_t \left[\barX_TF'\left(\barX_T / g(\mP^{t}_{\barX_T})  \right)\right]},\quad t\in[0,T].
\end{equation}
We are going to show that  (\ref{CRRAFOC_ZE}) is equivalent to (\ref{CRRAFOC}).
Indeed, by (\ref{DAansatzX}) and (\ref{DAansatzg}), we have, for $s\in[t,T]$, 
\begin{equation*}\label{CRRAFOC_computation}
	\begin{aligned}
		&\mE_s \left[\barX_T F'\left(\barX_T / g(\mP^{t}_{\barX_T})   \right)\right]\\
		&=\barX_s e^{\int_s^Ta_r^\mt\kappa(r)\md r-{1\over2}A(s)}\mE_s\left[e^{\int_s^Ta_r\md W_r}F’\left(e^{\int_t^Ta_r\md W_r}/H(A(t))\right)\right]\\
		&=\barX_s e^{\int_s^Ta_r^\mt\kappa(r)\md r-{1\over2}A(s)}
		\left(\mE\left[e^{\sqrt{A(s)}\xi}F’\left(e^{\sqrt{A(s)}\xi+y}/H(A(t))\right)\right]\right)_{y=\int_t^sa_r\md W_r},
	\end{aligned}
\end{equation*}
where $\xi\sim\N(0,1)$.
Consequently, 
\begin{align}
	&\mE_t\left[\barX_TF'\left(\barX_T/g(\mP^{t}_{\barX_T})\right)\right]
	=\barX_te^{\int_t^Ta_r^\mt\kappa(r)\md r-{1\over2}A(t)}\mE\left[e^{\sqrt{A(t)}\xi}F’\left(e^{\sqrt{A(t)}\xi}/H(A(t))\right)\right],  \label{CRRAEexpression} \\
	&Z^{\barX_TF'\left(\barX_T/g(\mP^{t}_{\barX_T})\right)}(t)\nonumber\\
	=&\barX_te^{\int_t^Ta_r^\mt\kappa(r)\md r-{1\over2}A(t)}\mE\left[e^{\sqrt{A(t)}\xi}\left(F’\left({e^{\sqrt{A(t)}\xi}\over H(A(t))}\right)+{e^{2\sqrt{A(t)}\xi}\over H(A(t))}F^{\prime\prime}\left({e^{\sqrt{A(t)}\xi}\over H(A(t))}\right)\right)\right]a_t.\label{CRRAZexpression}
\end{align}
Clearly, (\ref{CRRAFOC_ZE}) is equivalent to (\ref{CRRAFOC}), by (\ref{CRRAEexpression})--(\ref{CRRAZexpression}), which obviously implies  \eqref{ode:A}. 

\subsection{Proof of Theorem \ref{thm:crra}.} \label{CRRAproofapp} 
Let $A(t)=\cG^{-1}\left(\int_t^T|\kappa(s)|^2ds\right)$ and $a_t=\kappa(t)G(A(t))$ for all $t\in[0,T)$. Lemma \ref{lma:ode} implies that $A$ satisfies ODE \eqref{ode:A}. Therefore, $A(t)=-\int_t^T A^\prime(s)ds=\int_t^T|a_s|^2ds$, $t\in[0,T]$ and hence 
$\int_0^T|a_s|^2 \md s=A(0)<\infty$. Obviously, \eqref{eq:barpi:crra:G} implies  $\barpi_t=(\sigma^\mt (t))^{-1}a_t$ , hence $\barpi\in H^d_{\mathrm{BMO}}$.  To prove that $\barpi$ is a type-I equilibrium, we need to verify Conditions (a), (b) and (c) of Theorem \ref{verification-pc}. Condition (a) is obtained from Proposition \ref{CRRAeqprop}. We now verify Conditions (b) and (c).

    {\bf \underline{Verification of Condition (b)}.} We claim that, for fixed $t\in [0,T)$, $\barX q^t \in S^p(\mF,\mR^d)$ for any $p>1$.\footnote{Recall that in Section \ref{generalCRRA} we omit the superscripts $\Si$ and $\Oi$.} To prove this, by definitions of $p^t$ and $q^t$ (see Lemma \ref{adjoint}), we only need to show $\cX^t \in S^p(\mF,\mR)$ and $Z^{\barX_T\xi^t}\in S^p(\mF,\mR^d)$, where $\cX^t\defeq \{\mE_s[\barX_T\xi^t],s\in [0,T] \}$. From the proof of Proposition \ref{CRRAeqprop} we know that there exists a constant $C>0$ such that 
\[
0\leq \barX_T \xi^t \leq C\barX_T (R(t,T)^{\alpha_0}+R(t,T)^{-\alpha_0}).
\]
Recalling that $\barX_T=e^{\int_0^T a_s\cdot W_s+\int_0^T a_s^\mt \kappa(s)\md s -\frac{1}{2}A(0)}$ and $R(t,T)=e^{\int_t^T a_s\cdot \md W_s}$, we get $\cX^t\in S^p(\mF,\mR)$ directly from the properties of the Brownian motion. $Z^{\barX_T\xi^t}\in S^p(\mF,\mR^d)$ can be similarly proved.

{\bf \underline{Verification of Condition (c)}.}  We recall that
\[
	M_1(\mP^{t}_{\barX_T},\mP^{t}_{\barX_T^{t,\epsilon,\varphi}})=\frac{g(\mP^{t}_{\barX^{t,\epsilon,\varphi}_T})}{g(\mP^{t}_{\barX_T})}\cdot \frac{\mE_t \left[\barX_TF'\left(\frac{\barX_T}{g(\mP^{t}_{\barX_T})}    \right)\right]}{\mE_t\left[\barX^{t,\epsilon,\varphi}_TF'\left(\frac{\barX^{t,\epsilon,\varphi}_T}{g(\mP^{t}_{\barX_T})}    \right)\right]}.
	\]
	The intergrablity of $\barX^{t,\epsilon,\varphi}$ and the Dominated convergence theorem yield
	\[
	\lim_{\epsilon\to 0}\frac{\mE_t\left[ \barX_TF'\left(\frac{\barX_T}{g(\mP^{t}_{\barX_T})}    \right)\right]}{\mE_t\left[\barX^{t,\epsilon,\varphi}_TF'\left(\frac{\barX^{t,\epsilon,\varphi}_T}{g(\mP^{t}_{\barX_T})}    \right)\right]}=1, {\ \ \rm a.s.}.
	\]
On the other hand, in a similar way to \eqref{DAansatzg}, we can get
\[
g(\mP^t_{\barX_T^{t,\epsilon,\varphi}})=\barX_t e^{\int_t^{t+\epsilon}\varphi^\mt \theta(s)\md s+\int_t^T a_s^\mt \kappa(s)\md s -\frac{1}{2}\tilde{A}_\epsilon(t)}H(\tilde{A}_\epsilon(t)),
\]
where
\[
\tilde{A}_\epsilon(t) =\int_t^T|\tilde{a}^{t,\varphi,\epsilon}_s|^2\md s=\int_t^T |a_s + \sigma(s)^\mt\varphi\ind_{[t,t+\epsilon)}|^2\md s.
\]
By the continuity of $H$, we have $\lim_{\epsilon\to 0} g (\mP^t_{\barX_T^{t,\epsilon,\varphi}})=g(\mP^t_{\barX_T})$,  hence $\lim_{\epsilon\to 0} M_1(\mP^{t}_{\barX_T},\mP^{t}_{\barX_T^{t,\epsilon,\varphi}})=1$, proving Condition (c).

\section{Proofs and Calculations for Section \ref{WUrandom}.}

\subsection{Proof of Lemma \ref{lma:bsde}.}\label{sec:lma:bsde}
First, it easily follows that $(\hat{X},\hat{Y}_1,\hat{Y}_2,\hat{Z}_1,\hat{Z}_2)$ is the unique solution to the following FBSDEs,
\begin{align}	
	& \left\{
	\begin{aligned}
	&\md \hat{Y}_i(s)=-\frac{1}{2}(\hat{Z}_i(s))^2\md s+\hat{Z}_i(s) \md W_s,\ \ i=1,2,  \\ 	
  	&\hat{Y}_1(T)=r_1\hat{X}_T,\hat{Y}_2(T)=r_2\hat{X}_T,
    \end{aligned}
     \right.   \label{XY-BSDE-Y}   \\
    & \left\{
    \begin{aligned}
    	&\md \hat{X}_s=(\barpi_s\theta(s)-\frac{1}{2}\sigma(s)^2\barpi_s^2)\md s+\sigma(s)\barpi(s) \md W_s,\\ 	
    	&\hat{X}_0=\log x_0.
    \end{aligned}
    \right. \label{XY-BSDE-X}
\end{align}
Using (\ref{WU-derivative}), 
\begin{equation}\label{WU-xiXexpression}
\barX_T \xi^t=\barX_T \partial_x \nabla g(\mP^t_{\barX_T},\barX_T)=\frac{r_2\barX_T^{r_2}\mE_t[\barX^{r_1}]-r_1\barX_T^{r_1}\mE_t[\barX_T^{r_2} ]}{(1-\rho)(\mE_t [\barX_T^{r_1}])^2}.
\end{equation}
Then, 
\begin{equation}\label{WU-xiXexpression-random}
	\mE_s[\barX_T\xi^t]=\frac{\lambda_2Y_2(s)Y_1(t)+\lambda_1Y_1(s)Y_2(t)}{Y_1(t)^2}.
\end{equation}
Consequently,
\begin{align*}
	\mE_t[\barX_T\xi^t]=\frac{Y_2(t)}{Y_1(t)}\quad\text{and}\quad
  Z^{\barX_T\xi^t}(t)=\frac{\lambda_2Z_2(t)Y_1(t)+\lambda_1 Z_1(t)Y_2(t)}{Y_1(t)^2}.
\end{align*}
Direct calculations show that the first-order condition \eqref{FOC} yields
\begin{equation}\label{WUrelation}
\sigma(t) \barpi_t=\kappa(t)+\lambda_2\hat{Z}_2(t)+\lambda_1 \hat{Z}_1(t), \quad t\in [0,T].
\end{equation}
Thus, using \eqref{XY-BSDE-Y}--\eqref{XY-BSDE-X}, we know that $\barpi$ satisfies (\ref{FOC}) if and only if $(\hat{X},\hat{Y}_1,\hat{Y}_2,\hat{Z}_1,\hat{Z}_2)$ solves the following system of (coupled) quadratic FBSDEs:
\begin{align}
	&\left\{
	\begin{aligned}
		&\md \hat{Y}_i(s)=-\frac{1}{2}(\hat{Z}_i(s))^2\md s+\hat{Z}_i(s) \md W_s,\ \ i=1,2,\\ 	
		&\hat{Y}_1(T)=r_1 \hat{X}_T,\hat{Y}_2(T)=r_2\hat{X}_T,
	\end{aligned}
	\right. \label{WUFBSDE-B}\\
&\left\{
	\begin{aligned}
		\md \hat{X}_s=&[\kappa(s)+\lambda_2\hat{Z}_2(s)+\lambda_1 \hat{Z}_1(s)]\cdot\left[\left(\frac{1}{2}[\kappa(s)-(\lambda_2\hat{Z}_2(s)+\lambda_1\hat{Z}_1(s))] \right)\md s+ \md W_s\right],\\ 
		\hat{X}_0=&\log x_0.
	\end{aligned}
\right.\label{WUFBSDE-F}
\end{align}
Using the relation $\bar{Y}^i=\hat{Y}^i-r_i\hat{X}$, we immediately get the desired equivalence. 

\subsection{Calculations of the Coefficients of (\ref{WUQBSDE}).}\label{WUQBSDEcoefficient}
In this appendix we derive the explicit results of the coefficients $\bC^i$, $\bc_{i,j}$, and $\bfb_i$ in (\ref{WUQBSDE}).

Denote $\bar{\kappa}(s)=\left(\begin{array}{c} r_1\kappa(s) \\ r_2\kappa(s)   \end{array}\right)$. With this notation, the transform $\bar{Z}_i=(1-\lambda_ir_i)\hat{Z}_i(s)-r_i\lambda_{-i}\hat{Z}_{-i}-r_i\kappa(s)$ can be rewritten as $\bar{Z}=\bP\hat{Z}-\bar{\kappa}$, where
\[
\bP=\left( \begin{array}{cc}
	1-\lambda_1r_1   &  -r_1\lambda_2\\
	-r_2\lambda_1    &  1-\lambda_2r_2
\end{array}     \right)=I-\br \blambda^\mt,
\]
with $\br=(r_1, r_2)^\mt $, $\blambda=(\lambda_1 ,\lambda_2)^\mt$.
Therefore, $\hat{Z}=\bP^{-1}(\bar{Z}+\bar{\kappa})$. Recalling the definition of $r_i$ and $\lambda_i$ and by some calculations, we have  ${\rm det}(\bP)=\rho-2\gamma$ and 
\[
\bP^{-1}=\frac{1}{\rho-2\gamma}\left(\begin{array}{cc}
	1-\lambda_2r_2 & \lambda_2r_1\\
	\lambda_1r_2  & 1-\lambda_1r_1
\end{array}   \right)=\frac{1}{\rho-2\gamma}(I-\blambda_\bot\br_\bot^\mt),
\]
with $\br_\bot=(-r_2, r_1)^\mt$ and $\blambda_\bot=(-\lambda_2 ,\lambda_1)^\mt$.
On the other hand, the first equation in (\ref{WUQBSDE-1}) can be rewritten as follows:
\begin{equation}\label{WUQBSDE-2}
\md \bar{Y}_i(s)=-\frac{1}{2}[\hat{Z}(s)^\mt \bQ_i \hat{Z}(s)+r_i|\kappa(s)|^2]\md s+\bar{Z}(s)\md W_s,
\end{equation}
with
\[
\bQ_1=\left(\begin{array}{cc}
	1-r_1\lambda_1^2 & -r_1\lambda_1\lambda_2\\
	-r_1\lambda_1\lambda_2  & -r_1\lambda_2
\end{array}   \right),\ \ \bQ_2=\left(\begin{array}{cc}
-r_2\lambda_1^2 & -r_2\lambda_1\lambda_2\\
-r_2\lambda_1\lambda_2  & 1-r_2\lambda_2
\end{array}   \right).
\]
To summarize, $\bQ_i=E_i-r_i\blambda\blambda^\mt$, where $E_1=\left(\begin{array}{cc}
	1 & 0\\
	0 &   0
\end{array}   \right)$, $E_2=\left(\begin{array}{cc}
0 & 0\\
0 &   1
\end{array}   \right)$. 

Plugging the transformation $\hat{Z}=\bP^{-1}(\bar{Z}+\bar{\kappa})$ into (\ref{WUQBSDE-2}) and comparing it to (\ref{WUQBSDE}), we have,  for $i=1,2$, $\bC^i=(\bP^{-1})^\mt\bQ_i\bP^{-1}$, $\bc_i\defeq \left(\begin{array}{c}
\bc_{1,i} \\
\bc_{2,i}
\end{array}   \right)=2\bC^i \br$ and $b_i=\br^\mt \bC^i \br+r_i$. Therefore, 
\begin{align*}
	\bC^1=&(\bP^{-1})^\mt E_1 \bP^{-1}-r_1(\bP^{-1})^\mt\blambda\blambda^\mt \bP^{-1}\\
	=&\frac{1}{(\rho-2\gamma)^2} \left(\begin{array}{cc}
		(1-\lambda_2r_2)^2 & \lambda_2r_1(1-\lambda_2r_2)\\
		\lambda_2r_1(1-\lambda_2r_2)  & \lambda_2^2r_1^2
	\end{array}   \right)-\frac{r_1}{(\rho-2\gamma)^2}\blambda\blambda^\mt\\
   =&\frac{1}{(\rho-2\gamma)^2(1-\rho)^2}\times \\
   &\left(\begin{array}{cc}
   	\gamma^4+(3-4\rho)\gamma^3+(1-\rho)(4-6\rho)\gamma^2-4\rho(1-\rho)^2\gamma+\rho^2(1-\rho)^2& (1-\rho+\gamma)^2(\rho-\gamma)\gamma\\
   	(1-\rho+\gamma)^2(\rho-\gamma)\gamma  & \gamma(1-\rho+\gamma)^2(\gamma-1)
   \end{array}   \right).
\end{align*}
Here, we have just used the fact that $(\bP^{-1})^\mt \blambda =\frac{1}{\rho-2\gamma}\blambda$. Similarly, we have
\begin{align*}
\bC^2 =& \frac{1}{(\rho-2\gamma)^2(1-\rho)^2}\times\\
&\left(\begin{array}{cc}
	\gamma^2(1-\rho+\gamma)(-\rho+\gamma)& (1-\rho+\gamma)(1-\gamma)\gamma^2\\
	(1-\rho+\gamma)(1-\gamma)\gamma^2  & \gamma^4-\gamma^3-(1-\rho)\gamma^2-3(1-\rho)^2\gamma+(1-\rho)^2\rho
\end{array}   \right).
\end{align*}

Plugging the above results on $\bC^i$ into $\bc_i=2\bC^i\br$ yields
\begin{align*}
	\bc_1=& \frac{2}{(\rho-2\gamma)^2(1-\rho)}\left(\begin{array}{c}
    \gamma[\gamma^2-(1-\rho)\gamma+\rho(1-\rho)]\\
    	-\gamma(1-\rho+\gamma)^2  
    \end{array}   \right).\\
\bc_2=& \frac{2}{(\rho-2\gamma)^2(1-\rho)}\left(\begin{array}{c}
    \gamma[\gamma^2-(1-\rho)\gamma+\rho(1-\rho)]\\
    	-\gamma(1-\rho+\gamma)^2  
    \end{array}   \right).
\end{align*}
Therefore,
\begin{equation}\label{cmatrix}
\begin{aligned}
\mathbf{c}&\defeq \left(\begin{array}{cc} \bc_{1,1} & \bc_{1,2}\\ \bc_{2,1} & \bc_{2,2}   \end{array}   \right)\\
&=\frac{2}{(\rho-2\gamma)^2(1-\rho)}\left(  \begin{array}{cc}   
	\gamma[(\rho-\gamma)(1-\rho)+\gamma^2]  & \gamma^2(1-\rho+\gamma) \\
	-\gamma (1-\rho+\gamma)^2  &[(\rho-3\gamma)(1-\rho)-\gamma^2](1-\rho+\gamma)
	\end{array} \right).
\end{aligned}
\end{equation}

Finally, based on the aforementioned results, we obtain
\begin{align*}
	\bfb_1=\br^\mt \bC^1 \br+r_1
	   =\frac{\gamma(2\rho-3\gamma-1)}{(\rho-2\gamma)^2},\ \ \bfb_2=\br^\mt \bC^2 \br+r_2
	=\frac{(1-\rho+\gamma)(\rho-3\gamma)}{(\rho-2\gamma)^2}.
\end{align*}

\subsection{Proof of Lemma \ref{WUQBSDE-existence}.} \label{proofQBSDEexistence}
In this appendix we first consider a type of two-dimensional QBSDE, then apply the result to the concrete example in Section \ref{WUrandom}. To be specific, we consider the following system of QBSDE:
\begin{equation}\label{generalQBSDE}
	Y^i_t=-\int_t^T [Z(s)^\mt C^i Z(s)+\sum_{k=1,2}c_{k,i}Z^k(s)\kappa(s)+b_i |\kappa(s)|^2]\md s-\int_t^T Z^i(s)\md W_s,
\end{equation}
where for $i=1,2$, $C^i\in \mR^{2\times 2}$, $b_i\in \mR$, and $c=(c_{k,i})_{k,i=1,2}\in \mR^{2\times 2}$. Moreover, $Z=(Z^1, Z^2)^\mt$ is a column vector. To show the wellposedness of (\ref{generalQBSDE}), we need several notations and technical preparations. 

For some probability measure $\mP'$, and a $(\mP',\mF)$-martingale $M$, define
\[
\|M\|_{{\rm BMO}_p(\mP')}\defeq \sup_{\tau\in\cT_{[0,T]}}\left\| \left(\mE^{\mP'}_\tau [|M_T-M_\tau|^p ] \right)^{1/p}       \right\|_\infty,
\]
where $p=1,2$. We denote by BMO$_p(\mP')$ the space of $(\mP',\mF)$-martingales $M$ such that $\|M\|_{{\rm BMP}_p}<\infty$, and $H_{{\rm BMO}_p(\mP')}=\{z\in L^2(\mF',\mR): \forall(\mP',\mF)-{\rm Brownian\ Motion\  }W',\int_0^\cdot z(s)\md W'(s)\in {\rm BMO}_p(\mP')  \} $. We also define $\|z\|_{{\rm BMO}_p(\mP')} \defeq \left\| \int_0^\cdot z(s)\md W'(s)  \right\|_{{\rm BMO}_p(\mP')}$.  If $\mP'=\mP$, we omit the entry of probability measure. Under this convention, the notations in this appendix are consistent with those in Subsection \ref{notations}: $\|\pi\|_{\rm BMO}=\|\pi\|_{{\rm BMO}_2(\mP)}$. We first present equivalence results of the BMO norms.
\begin{lemma}\label{BMO12equivalent}
	There exists a universal constant $K>0$ such that, for any probability measure $\mP'$ and any $M\in {{\rm BMO_2}(\mP')}$, we have
	\[
     \|M\|_{{\rm BMO_1}(\mP')}\leq \|M\|_{{\rm BMO_2}(\mP')}\leq K\|M\|_{{\rm BMO_1}(\mP')}.
	\]
\end{lemma}
\proof{Proof.}
	See \citet[Corollary 2.1]{Kazamaki1994}.
\hfill \Halmos\endproof 

\begin{lemma}\label{BMOmeasurechange}
	Let $\alpha\in L^{\infty}(\mF,\mR^d)$ and define a probability measure $\mP^\alpha$ by $\frac{\md \mP^\alpha}{\md \mP}= \cZ^{W^\Si}_T(\alpha)$ (recall the notation of exponential martingale in Subsection \ref{notations}). Then $H^d_{{\rm BMO}_2}\subseteq H^d_{{\rm BMO}_2(\mP^\alpha)}$. Moreover,
	there exist constants $l=l(\|\alpha\|_{\infty})$ and $L=L(\|\alpha\|_{\infty})$, which depend only on the bound of $\alpha$, such that, for any $z\in H^d_{{\rm BMO}_2}$, 	\[
	l(\|\alpha\|_{\infty})\|z\|_{{\rm BMO}_2(\mP^\alpha)}\leq \|z\|_{{\rm BMO}_2}\leq L(\|\alpha\|_{\infty})\|z\|_{{\rm BMO}_2(\mP^\alpha)}.
	\]
\end{lemma}
\proof{Proof.}
	See \citet[Lemma A.1]{Herdegen2021}.
\hfill \Halmos\endproof

\vskip 5pt
Recall $\Theta\defeq \int_0^T |\kappa(s)|^2\md s\in L^{\infty}(\F_T)$. Let $\|\Theta\|_{{\rm BMO}_1(\mP')}\defeq \|\mE^{\mP’}_\cdot[\Theta] - \mE^{\mP’}[\Theta]    \|_{{\rm BMO}_1(\mP')}$. Recall the notation $V(\Theta)\defeq \sup_\tau \|\Theta-\mE_{\tau}[\Theta]\|_{\infty}$. Then $\|\Theta\|_{\rm BMO_1}\leq V(\Theta)$. We further have the following lemma. 
\begin{lemma}
	For any $\alpha\in L^\infty(\mF,\mR)$ and $\mP^\alpha$ defined as in Lemma \ref{BMOmeasurechange}, we have
	\[
	\|\Theta\|_{{\rm BMO}_1(\mP^\alpha)}\leq 2V(\Theta).
	\]
\end{lemma}
\proof{Proof.}
	By the definition of $\mP^\alpha$, for any stopping time $\tau$,
	\[
	\mE_\tau^{\mP^\alpha} [\Theta]=\frac{\mE_\tau [\cZ^W_T(\alpha)\Theta]}{\cZ^W_\tau(\alpha)}.
	\]
	Thus, 
	\begin{align*}
		|\Theta-\mE^{\mP^\alpha}_\tau[\Theta]|&=\left| \frac{\mE_\tau [\cZ^W_T(\alpha)\Theta]-\cZ^W_\tau(\alpha) \Theta}{\cZ^W_\tau(\alpha)}                  \right|\\
		&=\frac{\left|\mE_\tau\left[ \cZ^W_T(\alpha)(\Theta-\mE_\tau[\Theta])\right]+\mE_\tau[ \cZ^W_T(\alpha)]\mE_\tau[\Theta]-\cZ^W_\tau(\alpha)\Theta\right|}{\cZ^W_\tau(\alpha)}\\
		&\leq \frac{\mE_\tau\left[ \cZ^W_T(\alpha)|\Theta-\mE_\tau [\Theta]|\right]+\cZ^W_\tau(\alpha)|\Theta-\mE_\tau[\Theta]|}{\cZ^W_\tau(\alpha)}\\
		&\leq 2\sup_{\tau}\|\Theta-\mE_\tau [\Theta]\|_\infty\\
		&= 2V(\Theta),
	\end{align*}
which obviously leads to the desired result.
\hfill \Halmos\endproof 

\vskip 5pt
We now consider the QBSDE (\ref{generalQBSDE}). We first remark that the arguments in \cite{Tevzadze2008} or \cite{Kramkov2016} can not be  directly applied due to the presence of the {\it linear} term $\sum_k c_{k,i}Z^k(s)\kappa(s)$. Indeed, with this term, Assumption (A4) in the appendix of \cite{Kramkov2016} becomes 
\[
|f(t,u)-f(t,v)|\leq {\rm Const.}(|u-v|)(1+|u|+|v|).
\]
This in turn invalidates the proof of Lemma A.4 therein because the Lipshicz constant of the map $F$ can not be controlled even if $\|\zeta\|_{\rm BMO}$ is small; see also Remark 2.2 of \cite{Frei2014}. To overcome this, we first consider a version of (\ref{generalQBSDE}) whose linear term appears diagnolly, i.e., $c_{1,2}=c_{2,1}=0$. We use an inhomogeneous measure change and prove that the decoupled solution mapping is a contraction under suitable assumptions. 
\begin{lemma}\label{QBSDEdiaglinear}
	Define
	\[
	B=B(C^1,C^2,c_{1,1},c_{2,2},b,\|\kappa\|_{\infty})\defeq \max_{i=1,2}\left\{ 4KL(c_{i,i}\|\kappa\|_\infty)\left( \frac{\|C^i\|}{l(c_{i,i}\|\kappa\|_{\infty})^2} +|b_i|\right)  \right\}.
	\]
	Suppose $0<V(\Theta)<1/4B^2$ and $c_{1,2}=c_{2,1}=0$. Then (\ref{generalQBSDE}) admits a unique solution $(Y,Z)\in (L^{\infty}(\mF,\mR))^2\times H_{\rm BMO_2}^2$ 	with $\|Z\|_{\rm BMO_2}\leq R\defeq \frac{1-\sqrt{1-4B^2V(\Theta)}}{2B}$.
\end{lemma}

\proof{Proof.}
 Given $z\in H^2_{\rm BMO_2}$, we consider the map $\sF:z\mapsto Z$, where $Z=(Z^1,Z^2)^\mt$ is {given by the solution of }
 \begin{equation}\label{QBSDEdecoupled}
 Y^i_t=-\int_t^T (z(s)^\mt C^i z(s)+c_{i,i}Z^i(s)\kappa(s)+b_i|\kappa(s)|^2)\md s -\int_t^T Z^i(s)\md W(s).
 \end{equation}
 Clearly the map is well-defined because \eqref{QBSDEdecoupled} is a standard linear BSDE. Define a probability measure $\mP^i$ by the measure change $\frac{\md \mP^i}{\md \mP}=\cZ^W_T(-c_{i,i}\kappa)$. Based on Girsanov's theorem, $W^i\defeq W+c_{i,i}\int_0^\cdot \kappa (s)\md s$ is a $\mP^i$-Brownian motion. Under this measure change, (\ref{QBSDEdecoupled}) becomes
 \[
 Y^i_t=-\int_t^T z(s)^\mt C^i z(s)\md s-b_i \int_t^T |\kappa(s)|^2\md s-\int_t^T Z^i(s)\md W^i(s).
 \]
 Taking $\mE^{\mP^i}_t$ on both sides yields
 \[
 Y^i_t=-\mE_t^{\mP^i}\left[\int_t^T z(s)^\mt C^iz(s)\md s\right]-b_i\mE_t^{\mP^i}\left[\int_t^T|\kappa(s)|^2\md s\right].
 \]
 Therefore, for any stopping time $\tau$,
 \begin{equation}\label{Zequation}
 \int_\tau^T Z^i(s)\md W^i_s=\mE^{\mP^i}_\tau \left[\int_\tau^T z(s)^\mt C^iz(s)\md s\right]-\int_\tau^T z(s)^\mt C^iz(s)\md s+b_i(\mE^{\mP^i}_\tau [\Theta]-\Theta).
 \end{equation}
Taking absolute value and $\mE^{\mP^i}_\tau$, and recalling the definitions of various BMO norms, we derive from (\ref{Zequation}) and Lemma \ref{BMOmeasurechange} that 
\begin{align*}
	\|Z^i\|_{{\rm BMO}_1(\mP^i)}\leq 2\|C^i\|\|z\|^2_{{\rm BMO}_2(\mP^i)}+|b_i|\|\Theta\|_{{\rm BMO}_1(\mP^i)}  \leq\frac{2\|C^i\|}{l(c_{i,i}\|\kappa\|_{\infty})^2}\|z\|^2_{\rm BMO_2}+2|b_i|V(\Theta). 
\end{align*}
Using Lemmas \ref{BMO12equivalent} and \ref{BMOmeasurechange},
\[
\|Z^i\|_{\rm BMO_2}\leq 2KL(c_{i,i}\|\kappa\|_{\infty})\left(\frac{\|C^i\|}{l(c_{i,i}\|\kappa\|_{\infty})^2}\|z\|^2_{\rm BMO_2}+|b_i|V(\Theta)\right).
\]
Thus
\[
\|Z\|_{\rm BMO_2}\leq \|Z^1\|_{\rm BMO_2}+\|Z^2\|_{\rm BMO_2}\leq B(\|z\|^2_{\rm BMO_2}+V(\Theta)).
\]
Therefore, if $\|z\|_{\rm BMO_2}\leq R$, we have $\|\sF(z)\|_{\rm BMO_2}\leq B(R^2+V(\Theta))=R$. That is, $\sF$ maps $\B_R\defeq \{z\in H_{\rm BMO_2}:\|z\|_{\rm BMO_2}\leq R   \}$ to itself. Moreover, for $z,z'\in \B_{R}$, with $\delta z\defeq z-z'$, $\delta Z\defeq \sF(z)-\sF(z')$, we know from (\ref{Zequation}) that 
\[
\|\delta Z^i\|_{{\rm BMO}_1(\mP^i)}\leq 2\|C^i\|\|\delta z\|_{{\rm BMO}_2(\mP^i)}(\|z\|_{{\rm BMO}_2(\mP^i)}+\|z'\|_{{\rm BMO}_2(\mP^i)}).
\]
Using Lemmas \ref{BMO12equivalent} and \ref{BMOmeasurechange},
\[
\|\delta Z^i\|_{\rm BMO_2}\leq \frac{2\|C^i\|KL(c_{i,i}\|\kappa\|_{\infty})}{l(c_{i,i}\|\kappa\|_{\infty})^2}\|\delta z\|_{{\rm BMO}_2}(\|z\|_{{\rm BMO}_2}+\|z'\|_{{\rm BMO}_2})\leq BR\|\delta z\|_{{\rm BMO}_2}.
\]
Then, 
\[
\|\delta Z\|_{\rm BMO_2}\leq \|\delta Z^1\|_{\rm BMO_2}+\|\delta Z^2\|_{\rm BMO_2}\leq 2BR\|\delta z\|_{\rm BMO_2}.
\]
Noting that $R< 1/2B$, we conclude that $\sF$ is a contraction on $\B_{R}$, thus the desired result follows by contraction mapping principle.
\hfill \Halmos\endproof 

\vskip 5pt

We now provide the following proposition.
\begin{proposition}\label{QBSDEsolvability}
	Suppose that the matrix $c=(c_{k,i})_{k,i=1,2}$ has two real eigenvalues. Then for any sufficiently small $\varrho>0$, there exists $V_0>0$ such that if $V(\Theta)<V_0$, the QBSDE (\ref{generalQBSDE}) admits a unique solution $(Y,Z)\in (L^{\infty}(\mF,\mR))^2\times H^2_{\rm BMO_2}$ with $\|Z\|_{\rm BMO_2}<\varrho$.
\end{proposition}
\proof{Proof.}
	Suppose that $\lambda_1,\lambda_2\in \mR$ are two eigenvalues of $c$ (possibly the same), with $u=(u_1,u_2)^\mt$, $v=(v_1,v_2)^\mt$ their eigenvectors, respectively. We can also assume that $u$ and $v$ are linear independent, thus $P\defeq\left(\begin{array}{cc}u_1 &u_2 \\ v_1 &v_2    \end{array}    \right)$ is invertible.	Now let us consider the linear transform $\cZ\defeq PZ, \cY\defeq PY$. Then $(Y,Z)$ solves (\ref{generalQBSDE}) if and only if $(\cY,\cZ)$ solves
	\begin{equation}\label{QBSDEtransform}
		\left\{
		\begin{aligned}
			&\cY^1_t=-\int_t^T[\cZ(s)^\mt (P^\mt)^{-1}(u_1C^1+u_2C^2)P^{-1}\cZ(s)+\lambda_1 \cZ^1(s)+u^\mt b |\kappa(s)|^2]\md s-\int_t^T\cZ^1(s)\md W_s,\\
			&\cY^2_t=-\int_t^T[\cZ(s)^\mt (P^\mt)^{-1}(v_1C^1+v_2C^2)P^{-1}\cZ(s)+\lambda_2 \cZ^2(s)+v^\mt b |\kappa(s)|^2]\md s-\int_t^T\cZ^2(s)\md W_s.
		\end{aligned}
		\right.
	\end{equation} 
Here, we have  used that $cu=\lambda_1u$, $cv=\lambda_2v$. Note that Lemma \ref{QBSDEdiaglinear} is now applicable to prove the existence of solution to (\ref{QBSDEtransform}), provided that $V(\Theta)$ is sufficiently small. This completes the proof.
\hfill \Halmos\endproof 

\proof{Proof of Lemma \ref{WUQBSDE-existence}.}
Because $-1<\gamma \leq 0$, $\gamma\leq \rho<1+\gamma$, it is clear that $\gamma[(\rho-\gamma)(1-\rho)+\gamma^2]<0$ and $-\gamma^3(1-\rho+\gamma)^3>0$. Moreover, $[(\rho-3\gamma)(1-\rho)-\gamma^2](1-\rho+\gamma)>-\gamma(2(1-\rho)+\gamma)(1-\rho+\gamma)\geq \gamma^2(1-\rho+\gamma)\geq 0$. Therefore, by \eqref{cmatrix}, ${\rm det}(\mathbf{c})<0$ and thus $\mathbf{c}$ has two real eigenvalues. Then, as a direct application of Proposition \ref{QBSDEsolvability}, we obtain the desired result.
\hfill \Halmos\endproof 

\subsection{Proof of Proposition \ref{verificationWU-random}.}\label{proofverificationWU}
In this subsection we need estimations of {\it exponential processes}, studied in \cite{Yong2006}. Defining $M(t;r_1,r_2)\defeq \exp\left(-\int_0^s\left[r_1(s)+\frac{1}{2}|r_2(s)|^2    \right] \md s - \int_0^t r_2(s)\md W_s      \right)$, we have $\E^\varphi=M(\cdot;-\varphi^\mt \theta+\varphi^\mt \sigma \sigma^\mt \barpi,-\varphi^\mt \sigma)$. The integrability of the process $M(\cdot;r_1,r_2)$ mainly lies in the {\it exponential integrability} of $r_1$ and $r_2$. We list here the main results we will use. 
\begin{lemma}\label{Yongthmexplicit}
	For $\alpha>0$, $\beta>1$, we have
	\begin{equation}\label{expestimate}
		\begin{aligned}
			&\mE\left[\sup_{t\in [0,T]}M(t;r_1,r_2)^{\frac{\alpha\beta}{\beta+\alpha(2\sqrt{\beta}-1)}}    \right] \\
			&\ \ \ \leq \left\{ \mE\left[e^{\frac{\beta}{2}\int_0^T|r_2(s)|^2\md s}  \right]    \right\}^{\frac{\alpha(\sqrt{\beta}-1)}{\beta+\alpha(2\sqrt{\beta}-1)}}	\cdot \left\{\mE\left[ \sup_{t\in[0,T]}e^{-\alpha\int_0^t r_1(s)\md s}   \right]      \right\}^{\frac{\beta}{\beta+\alpha(2\sqrt{\beta}-1)}}.
		\end{aligned}
	\end{equation}
\end{lemma}
\proof{Proof.}
	See \citet[Theorems 3.4 and 3.5]{Yong2006}.
\hfill \Halmos\endproof 
\begin{lemma}\label{Yongthm}
	If, for some $\alpha_1,\alpha_2>0$ and $\beta>1$, we have
	\begin{align}
		&\mE\left[\sup_{s\in[0,T]}\left(e^{-\alpha_1 \int_0^t r_1(s)\md s}+e^{\alpha_2 \int_0^t r_1(s)\md s}\right)\right]<\infty,\label{r1}\\ &\mE\left[e^{\frac{\beta}{2}\int_0^T |r_2(s)|^2\md s}\right]<\infty,\label{r2}
	\end{align}
	then
	\[
	\mE\left[\sup_{s\in [0,T]}\left(M(s;r_1,r_2)^{\frac{\alpha_1\beta}{\beta+\alpha_1(2\sqrt{\beta}-1)}}+M(s;r_1,r_2)^{-\frac{\alpha_2\beta}{\beta+\alpha_2(2\sqrt{\beta}+1)}} \right)\right]<\infty.
	\]
	\proof{Proof.}
		This is a direct consequence of Lemma \ref{Yongthmexplicit}.
	\hfill \Halmos\endproof 
\end{lemma}
\vskip 5pt
\proof{Proof of Proposition \ref{verificationWU-random}.}
 Using the energy inequality of BMO norms (see, e.g. Lemma C.1 of \cite{Fu2023}) and Expression (\ref{WUrelationtransform}), we know that there exists a constant $K=K(\rho,\gamma)$ (only depending on $\rho$ and $\gamma$) such that, for any $0<c<\frac{1}{K\|\bar{Z}\|^2_{\rm BMO}}$, we have
	\[
	\mE \left[e^{c\int_0^T |\barpi(s)|^2\md s}\right]<\infty.
	\]
	Combining this observation with Lemma \ref{Yongthm}, we know that, assuming $V(\Theta)$ is sufficiently small such that $\|\bar{Z}\|_{\rm BMO}$ is also sufficiently small due to Lemma \ref{WUQBSDE-existence}, we can assume $\mE\left[\sup_{s\in [0,T]}(\barX_s^9+\barX_s^{-9})\right]<\infty$. Clearly, Assumption \ref{gassumption1} and Condition (c') of Theorem \ref{verification-pc} hold true if we can verify (\ref{Mintegral}). 
	
	Indeed, using Jensen's inequality and the fact $\mE_s[\barX_T^{r_i}]=\barX_s^{r_i}e^{\bar{Y}_i(s)}$, we have, for any $n\geq 1$, there exists a $K_n>0$ such that
	\begin{align*}
		\mE\left[ |M_1(\mP^{t}_{\barX_T^{t,\epsilon,\varphi}},\mP^{t}_{\barX_T})|^n\right] &\leq C\mE\left[\left(\frac{\barX_t^\gamma(\E^\varphi_t)^\gamma}{\mE_t\barX_{t+\epsilon}^\gamma(\E^\varphi_{t+\epsilon})^\gamma} \right)^n\right]\leq C\mE \left[\left(\mE_t\frac{\barX_t^\gamma(\E^\varphi_t)^\gamma}{\barX_{t+\epsilon}^\gamma(\E^\varphi_{t+\epsilon})^\gamma} \right)^n\right]\\
		&\leq C\mE\left[ \left( \frac{\barX_{t+\epsilon}\E^\varphi_{t+\epsilon}}{\barX_{t}\E^\varphi_{t}}    \right)^{|\gamma|n}\right]\leq C\mE \left[e^{K_n \int_t^{t+\epsilon}|\barpi_s|^2\md s}\right].
	\end{align*} 
	  The monotone convergence theorem yields
	\[
	\lim_{\epsilon\to 0}\mE \left[e^{K_n \int_t^{t+\epsilon}|\barpi_s|^2\md s}\right]=1.
	\]
	Consequently,
	\[
	\limsup_{\epsilon\to 0}\mE\left[ |M_1(\mP^{t}_{\barX_T^{t,\epsilon,\varphi}},\mP^{t}_{\barX_T})|^n\right]\leq 1,
	\]
	i.e., (\ref{Mintegral}) is verified, thus so is Condition (c'). 
	
	Using Lemma \ref{adjoint} and Expressions (\ref{WU-xiXexpression-random})-(\ref{WUrelation}), we have
	\begin{align*}
	(\kappa(s)-\sigma(s)^\mt\barpi_s)\mE_s[\barX_T \xi^t]+Z^{\barX_T\xi^t}(s)=&\left\{-\left[\lambda_1\hat{Z}_1(s)+\lambda_2\hat{Z}_2(s)\right]\right\}\cdot\frac{\lambda_1Y_1(s)Y_2(t)+\lambda_2Y_2(s)Y_1(t)}{Y_1(t)^2}\\
		&+\frac{\lambda_1\hat{Z}_1(s)Y_1(s)Y_2(t)+\lambda_2\hat{Z}_2(s)Y_2(s)Y_1(t)}{Y_1(t)^2}.
	\end{align*}
	Therefore, to verify Condition (b'), we only need to prove that for a.e. $t\in [0,T)$, there exist $m>1$ and $K_t>0$ such that for sufficiently small $\epsilon$,
	\[
	\mE\left[\sup_{t\leq s\leq t+\epsilon}\left(\frac{Y_i(s)|Y_j(t)-Y_j(s)|}{Y_1(t)^2}   \right)^m\right]\leq K_t \epsilon^{m/2},
	\]
	where $i,j=1,2$. Indeed, because $Y_i(s)=\barX_s^{r_i}e^{\bar{Y}_i(s)}$, $\hat{Z}_i\in H^d_{\rm BMO}$, and $\bar{Y}_i\in L^{\infty}(\mF,\mR)$, we have
	\begin{align*}
		\mE\left[\int_0^T |\hat{Z}_j(s)Y_j(s)|^2\md s\right]&\leq C\mE\left[\sup_s\barX_s^{2r_j}\int_0^T |\hat{Z}_i(s)|^2\md s\right]\\
		&\leq C \left(\mE\left[\sup_s \barX^2_s\right]\right)^{|r_j|}\left(\mE\left[\left(\int_0^T|\hat{Z}_i(s)|^2\md s \right)^{\frac{1}{1-|r_j|}} \right] \right)^{1-|r_j|}\\
		&<\infty.
	\end{align*}
	Moreover, by Lebesgue's differential theorem, for a.e. $t\in [0,T)$, we have
	\[
	\lim_{\epsilon\to 0}\frac{1}{\epsilon}\int_t^{t+\epsilon}\mE[|\hat{Z}_j(s)Y_j(s)|^2]\md s=\mE[|\hat{Z}_j(t)Y_j(t)|^2]<\infty.
	\] 
	As a consequence, for a.e. $t\in [0,T)$, there exists a $K'_t>0$ such that for sufficiently small $\epsilon>0$,
	\[
	\mE\left[\int_t^{t+\epsilon}|\hat{Z}_j(s)Y_j(s)|^2\md s\right]\leq K'_t\epsilon.
	\]
	As $1<m<18/17$ (so that $\frac{8m}{2-m}<9$), 
	\begin{align*}
		\mE\left[\sup_{t\leq s\leq t+\epsilon}\left(  \frac{Y_i(s)}{Y_1(t)^2}\right)^{\frac{2m}{2-m}}\right]&\leq \left\{\mE\left[\sup_{t\leq s\leq t+\epsilon}Y_i(s)^{\frac{4m}{2-m}}\right]    \right\}^{1/2}\left\{\mE \left[Y_1(t)^{-\frac{8m}{2-m}}\right] \right\}^{1/2} \\
		&\leq C\left\{\mE\left[\sup_{t\leq s\leq t+\epsilon}\barX_s^{\frac{4mr_i}{2-m}}\right]    \right\}^{1/2}\left\{\mE\left[ \barX_t^{|\gamma|\frac{8m}{2-m}}\right] \right\}^{1/2}\\
		&\leq C\left(\mE\left[\sup_{s\in [0,T]}\barX_s^9\right]+\mE\left[\sup_{s\in[0,T]}\barX_s^{-9}\right]\right)^{1/m'}\\
		&<\infty,
	\end{align*}
	for some $m'>1$. Then,
	\begin{align*}
		\mE\left[\sup_{t\leq s\leq t+\epsilon}\left(\frac{Y_i(s)|Y_j(t)-Y_j(s)|}{Y_1(t)^2}   \right)^m\right]&\leq \left\{\mE \left[\sup_{t\leq s\leq t+\epsilon}\left(\frac{Y_i(s)}{Y_1(t)^2}\right)^{\frac{2m}{2-m}}\right] \right\}^{1-\frac{m}{2}} \left\{\mE\left[\sup_{t\leq s\leq t+\epsilon}|Y_j(t)-Y_j(s)|^2\right]\right\}^{m/2}\\
		&\leq \left\{\mE \left[\sup_{t\leq s\leq t+\epsilon}\left(\frac{Y_i(s)}{Y_1(t)^2}\right)^{\frac{2m}{2-m}}\right] \right\}^{1-\frac{m}{2}}\left\{\mE\left[\int_t^{t+\epsilon}|\hat{Z}_j(s)Y_j(s)|^2\md s \right] \right\}^{m/2}\\
		&\leq K_t\epsilon^{m/2}.
	\end{align*}
	Here, we have  used It$\hato$'s formula to $Y_j$, which satisfies 
	\[
	\md Y_j(s)=Z_i(s)\md W_s=\hat{Z}_j(s)Y_j(s)\md W_s.
	\]
	
	Finally, we verify Condition (a) of Theorem \ref{verification-pc}. On the one hand, based on  the relation between $\hat{Z}_i$ and $\bar{Z}_i$, $i=1,2$, (\ref{FOC}) is obtained from Lemma \ref{lma:bsde}, and \eqref{WUrelationtransform} follows from \eqref{WUrelation}. On the other hand, note that in the current setting we have (recall (\ref{WU-xiXexpression}) and the definition of $(Y_1(t),Y_2(t))$):
	\[
	\barX_T\xi^t=\lambda_1 \barX_T^\gamma Y_2(t)+\lambda_2 \barX_T^{1-\rho+\gamma}Y_1(t). 
	\]
	For $s\in [0,T]$, $Y_i(s)=\barX_s^{r_i}e^{\bar{Y}_i(s)}$ with $\bar{Y}_1,\bar{Y}_2\in L^{\infty}(\mF,\mR)$. Thus, it is not hard to show $\barX_T \xi^t\in L^p(\F_T)$ for some $p>1$. To conclude, Condition (a) of Theorem \ref{verification-pc} is verified and the proof follows.
\hfill \Halmos\endproof 
\end{APPENDIX}

\bibliographystyle{abbrvnat}  
\bibliography{ref}

\begin{thebibliography}{39}
\providecommand{\natexlab}[1]{#1}
\providecommand{\url}[1]{\texttt{#1}}
\expandafter\ifx\csname urlstyle\endcsname\relax
  \providecommand{\doi}[1]{doi: #1}\else
  \providecommand{\doi}{doi: \begingroup \urlstyle{rm}\Url}\fi

\bibitem[Allais(1953)]{Allais1953}
M.~Allais.
\newblock {Le Comportement de l'Homme Rationnel devant le Risque: Critique des
  Postulats et Axiomes de l'Ecole Americaine}.
\newblock \emph{Econometrica}, 21\penalty0 (4):\penalty0 503--546, 1953.

\bibitem[Back(2017)]{Back2017}
K.~E. Back.
\newblock \emph{{Asset Pricing and Portfolio Choice Theory}}.
\newblock Oxford University Press, 2017.

\bibitem[Basak and Chabakauri(2010)]{Basak2010}
S.~Basak and G.~Chabakauri.
\newblock {Dynamic Mean-Variance Asset Allocation}.
\newblock \emph{The Review of Financial Studies}, 23\penalty0 (8):\penalty0
  2970--3016, 2010.

\bibitem[Bj{\"o}rk and Murgoci(2014)]{BM2014}
T.~Bj{\"o}rk and A.~Murgoci.
\newblock A theory of markovian time-inconsistent stochastic control in
  discrete time.
\newblock \emph{Finance and Stochastics}, 18:\penalty0 545--592, 2014.

\bibitem[Bj{\"{o}}rk et~al.(2014)Bj{\"{o}}rk, Murgoci, and Zhou]{Bjork2014}
T.~Bj{\"{o}}rk, A.~Murgoci, and X.~Y. Zhou.
\newblock {Mean-variance portfolio optimization with state-dependent risk
  aversion}.
\newblock \emph{Mathematical Finance}, 24\penalty0 (1):\penalty0 1--24, 2014.

\bibitem[Bj{\"o}rk et~al.(2017)Bj{\"o}rk, Khapko, and Murgoci]{Bjork2017}
T.~Bj{\"o}rk, M.~Khapko, and A.~Murgoci.
\newblock On time-inconsistent stochastic control in continuous time.
\newblock \emph{Finance and Stochastics}, 21\penalty0 (2):\penalty0 331--360,
  2017.

\bibitem[Carmona and Delarue(2018)]{Carmona2018}
R.~Carmona and F.~Delarue.
\newblock \emph{{Probabilistic Theory of Mean Field Games with Applications I:
  Mean Field FBSDEs, Control, and Games}}.
\newblock Springer Cham, 2018.

\bibitem[Chew(1983)]{Chew1983}
S.~H. Chew.
\newblock {A Generalization of the Quasilinear Mean with Applications to the
  Measurement of Income Inequality and Decision Theory Resolving the Allais
  Paradox}.
\newblock \emph{Econometrica}, 51\penalty0 (4):\penalty0 1065, 1983.

\bibitem[Chew(1989)]{Chew1989}
S.~H. Chew.
\newblock {Axiomatic utility theories with the betweenness property}.
\newblock \emph{Annals of Operations Research}, 19:\penalty0 273--298, 1989.

\bibitem[Dai et~al.(2021)Dai, Jin, Kou, and Xu]{Dai2021}
M.~Dai, H.~Jin, S.~Kou, and Y.~Xu.
\newblock {A dynamic mean-variance analysis for log returns}.
\newblock \emph{Management Science}, 67\penalty0 (2):\penalty0 1093--1108,
  2021.

\bibitem[Dai et~al.(2023)Dai, Dong, and Jia]{Dai2023}
M.~Dai, Y.~Dong, and Y.~Jia.
\newblock {Learning equilibrium mean-variance strategy}.
\newblock \emph{Mathematical Finance}, pages 1--47, 2023.

\bibitem[Dekel(1986)]{Dekel1986}
E.~Dekel.
\newblock {An axiomatic characterization of preferences under uncertainty:
  Weakening the independence axiom}.
\newblock \emph{Journal of Economic Theory}, 40\penalty0 (2):\penalty0
  304--318, 1986.

\bibitem[Ekeland and Lazrak(2006)]{Ekeland2006}
I.~Ekeland and A.~Lazrak.
\newblock Being serious about non-commitment: subgame perfect equilibrium in
  continuous time.
\newblock arXiv math/0604264, 2006.

\bibitem[Fan et~al.(2023)Fan, Hu, and Tang]{Fan2023}
S.~Fan, Y.~Hu, and S.~Tang.
\newblock Multi-dimensional backward stochastic differential equations of
  diagonally quadratic generators: The general result.
\newblock \emph{Journal of Differential Equations}, 368:\penalty0 105--140,
  2023.

\bibitem[Frei(2014)]{Frei2014}
C.~Frei.
\newblock {Splitting multidimensional BSDEs and finding local equilibria}.
\newblock \emph{Stochastic Processes and their Applications}, 124\penalty0
  (8):\penalty0 2654--2671, 2014.

\bibitem[Fu and Zhou(2023)]{Fu2023}
G.~Fu and C.~Zhou.
\newblock {Mean field portfolio games}.
\newblock \emph{Finance and Stochastics}, 27\penalty0 (1):\penalty0 189--231,
  2023.

\bibitem[Hamaguchi(2021{\natexlab{a}})]{Hamaguchi2021}
Y.~Hamaguchi.
\newblock {Time-inconsistent consumption-investment problems in incomplete
  markets under general discount functions}.
\newblock \emph{SIAM Journal on Control and Optimization}, 59\penalty0
  (3):\penalty0 2121--2146, 2021{\natexlab{a}}.

\bibitem[Hamaguchi(2021{\natexlab{b}})]{Hamaguchi2021a}
Y.~Hamaguchi.
\newblock {Extended backward stochastic volterra integral equations and their
  applications to time-inconsistent stochastic recursive control problems}.
\newblock \emph{Mathematical Control and Related Fields}, 11\penalty0
  (2):\penalty0 433--478, 2021{\natexlab{b}}.

\bibitem[Herdegen et~al.(2021)Herdegen, Muhle-Karbe, and
  Possama{\"{i}}]{Herdegen2021}
M.~Herdegen, J.~Muhle-Karbe, and D.~Possama{\"{i}}.
\newblock {Equilibrium asset pricing with transaction costs}.
\newblock \emph{Finance and Stochastics}, 25\penalty0 (2):\penalty0 231--275,
  2021.

\bibitem[Hern{\'{a}}ndez and Possama{\"{i}}(2023)]{Hernandez2023}
C.~Hern{\'{a}}ndez and D.~Possama{\"{i}}.
\newblock {Me, Myself and I: A general theory of non-Markovian
  time-inconsistent stochastic control for sophisticated agents}.
\newblock \emph{Annals of Applied Probability}, 33\penalty0 (2):\penalty0
  1196--1258, 2023.

\bibitem[Hu and Tang(2016)]{Hu2016}
Y.~Hu and S.~Tang.
\newblock Multi-dimensional backward stochastic differential equations of
  diagonally quadratic generators.
\newblock \emph{Stochastic Processes and their Applications}, 126\penalty0
  (4):\penalty0 1066--1086, 2016.

\bibitem[Hu et~al.(2012)Hu, Jin, and Zhou]{Hu2012}
Y.~Hu, H.~Jin, and X.~Y. Zhou.
\newblock Time-inconsistent stochastic linear-quadratic control.
\newblock \emph{SIAM journal on Control and Optimization}, 50\penalty0
  (3):\penalty0 1548--1572, 2012.

\bibitem[Hu et~al.(2017)Hu, Jin, and Zhou]{Hu2017}
Y.~Hu, H.~Jin, and X.~Y. Zhou.
\newblock Time-inconsistent stochastic linear-quadratic control:
  Characterization and uniqueness of equilibrium.
\newblock \emph{SIAM Journal on Control and Optimization}, 55\penalty0
  (2):\penalty0 1261--1279, 2017.
\newblock \doi{10.1137/15M1019040}.

\bibitem[Hu et~al.(2021)Hu, Jin, and Zhou]{Hu2021}
Y.~Hu, H.~Jin, and X.~Y. Zhou.
\newblock {Consistent investment of sophisticated rank-dependent utility agents
  in continuous time}.
\newblock \emph{Mathematical Finance}, 31\penalty0 (3):\penalty0 1056--1095,
  2021.

\bibitem[Jackson and \v{Z}itkovi\'{c}(2022)]{Jackson2022}
J.~Jackson and G.~\v{Z}itkovi\'{c}.
\newblock Existence and uniqueness for non-markovian triangular quadratic
  bsdes.
\newblock \emph{SIAM Journal on Control and Optimization}, 60\penalty0
  (3):\penalty0 1642--1666, 2022.

\bibitem[Jamneshan et~al.(2017)Jamneshan, Kupper, and Luo]{Jamneshan2017}
A.~Jamneshan, M.~Kupper, and P.~Luo.
\newblock {Multidimensional quadratic BSDEs with separated generators}.
\newblock arXiv 1501.00461v4, 2017.

\bibitem[Kazamaki(1994)]{Kazamaki1994}
N.~Kazamaki.
\newblock \emph{{Continuous Exponential Martingales and BMO}}.
\newblock Springer, Berlin, 1994.

\bibitem[Kramkov and Pulido(2016)]{Kramkov2016}
D.~Kramkov and S.~Pulido.
\newblock {A system of quadratic BSDEs arising in a price impact model}.
\newblock \emph{Annals of Applied Probability}, 26\penalty0 (2):\penalty0
  794--817, 2016.

\bibitem[Li and Ng(2000)]{Li2000}
D.~Li and W.-L. Ng.
\newblock Optimal dynamic portfolio selection: Multiperiod mean-variance
  formulation.
\newblock \emph{Mathematical Finance}, 10\penalty0 (3):\penalty0 387--406,
  2000.

\bibitem[Markowitz(1952)]{Markowitz1952}
H.~Markowitz.
\newblock Portfolio selection.
\newblock \emph{The Journal of Finance}, 7\penalty0 (1):\penalty0 77--91, 1952.

\bibitem[Merton(1969)]{Merton1969}
R.~C. Merton.
\newblock Lifetime portfolio selection under uncertainty: The continuous-time
  case.
\newblock \emph{The Review of Economics and Statistics}, 51\penalty0
  (3):\penalty0 247--257, 1969.

\bibitem[Merton(1971)]{Merton1971}
R.~C. Merton.
\newblock Optimum consumption and portfolio rules in a continuous-time model.
\newblock \emph{Journal of Economic Theory}, 3\penalty0 (4):\penalty0 373--413,
  1971.

\bibitem[Quiggin(1982)]{Quiggin1982}
J.~Quiggin.
\newblock A theory of anticipated utility.
\newblock \emph{Journal of Economic Behavior \& Organization}, 3\penalty0
  (4):\penalty0 323--343, 1982.

\bibitem[Quiggin(1993)]{Quiggin1993}
J.~Quiggin.
\newblock \emph{Generalized Expected Utility Theory: The Rank-Dependent Model}.
\newblock Kluwer, Boston, 1993.

\bibitem[Strotz(1955)]{Strotz1955}
R.~H. Strotz.
\newblock Myopia and inconsistency in dynamic utility maximization.
\newblock \emph{The Review of Economic Studies}, 23\penalty0 (3):\penalty0
  165--180, 1955.

\bibitem[Tevzadze(2008)]{Tevzadze2008}
R.~Tevzadze.
\newblock {Solvability of backward stochastic differential equations with
  quadratic growth}.
\newblock \emph{Stochastic Processes and their Applications}, 118\penalty0
  (3):\penalty0 503--515, 2008.

\bibitem[Yong(2006)]{Yong2006}
J.~Yong.
\newblock {Completeness of security markets and solvability of linear backward
  stochastic differential equations}.
\newblock \emph{Journal of Mathematical Analysis and Applications},
  319\penalty0 (1):\penalty0 333--356, 2006.

\bibitem[Zhang(2017)]{Zhang2017}
J.~Zhang.
\newblock \emph{{Backward Stochastic Differential Equations: From Linear to
  Fully Nonlinear Theory}}.
\newblock Springer New York, 2017.

\bibitem[Zhou and Li(2000)]{Zhou2000}
X.~Y. Zhou and D.~Li.
\newblock Continuous-time mean-variance portfolio selection: A stochastic lq
  framework.
\newblock \emph{Applied Mathematics and Optimization}, 42\penalty0
  (1):\penalty0 19--33, 2000.

\end{thebibliography}

\newpage
\setcounter{equation}{0} 
\renewcommand{\theequation}{OC. \arabic{equation}}

\begin{center}
 \vspace*{0pt}%
  \TITLEfont\HD{24}{0}The Online Companion \HD{0}{15}
\end{center}
\vskip 20pt

In this Online Companion, we apply Condition (a) in Theorem 3.2 of the main context to recover some results of %derive candidate equilibrium strategies for the problems %in 
\cite{Basak2010} and \cite{Hu2021} on portfolio selections for the mean-variance (MV) preference with stochastic {factor} models and for the rank dependent utility (RDU) with the Black-Scholes model, respectively.

\section*{OC1 Mean variance preference.}\label{MVexm}

Let us now investigate the dynamic MV problems. For simplicity, the interest rate is 0 and there is only one stock and only one stochastic factor; the extension  to the case of multiple stocks and stochastic factors is straightforward.  The stock price process $S$ and the stochastic factor $Y$ satisfy the following system of SDEs:
\[
\begin{aligned}
	&\md S_t/S_t = \theta(t,S_t,Y_t)\md t + \sigma(t,S_t,Y_t)\md W^\Si_t,\\
	&\md Y_t=m(t,Y_t)\md t+\nu(t,Y_t)\md W^Y_t,\\
	&W^Y=\rho W^\Si+\sqrt{1-\rho^2}W^\Oi,
\end{aligned}
\]
where $\rho\in[-1,1]$, Brownian motion $W^\Si$ is independent of Brownian motion $W^\Oi$, and all of $\theta$, $\sigma$, $m$ and $\nu$ are Borel functions such that the above system of SDEs has a unique  solution.  

The MV preference functional $g$ is given by
\[
g(\mP_X)=\mE[X]-\frac{\gamma}{2}\mE[X^2]+\frac{\gamma}{2}(\mE[X])^2,
\]
where $\gamma>0$ is the coefficient of risk aversion. 
Moreover, $\cP_0=\{\mu\in\cP(\mR): \int_\mR x^2\mu(dx)<\infty\}$.

It is straightforward to obtain
\[
\partial_x \nabla g(\mP_X,x)=1-\gamma x+\gamma \mE[X].
\]
Taking $\xi^t=\partial_x \nabla g(\mP^{t}_{\barX_T},\barX_T)$, (3.8) turns into a very simple form:
\[
Z^{\barX_T,\Si}(t)=\frac{\kappa(t)}{\gamma}.
\]
That is, there exist a process $\phi$ such that
\[
\barX_T=\mE[\barX_T]+\int_0^T\frac{\kappa(s)}{\gamma}\md W^{\Si}_s+\int_0^T \phi_s \md W^{\Oi}_s.
\]
Now the problem is to replicate $\barX_T$, i.e., to find appropriate $\pi$ such that the corresponding terminal wealth is $\barX_T$. Note that for $H_{\cdot}=\exp(-\frac{1}{2}\int_0^\cdot |\kappa_s|^2 \md s-\int_0^\cdot \kappa_s \md W^{\Si}_s)$ (only $W^{\Si}$ is involved), $H_\cdot \barX_\cdot$ is always a local martingale, we aim to find a $\pi$ such that it is a martingale. Therefore,
\begin{align*}
	X_t&=\frac{1}{H_t}\left( \mE_t\left[ H_T \int_0^T\frac{\kappa(s)}{\gamma}\md W^{\Si}_s  \right] +  \mE_t\left[ H_T \int_0^T\phi_s\md W^\Oi_s \right] \right)\\
	&=\int_0^t \frac{\kappa(s)}{\gamma}\md W^\Si_s +\int_0^t \phi_s \md W^\Oi_s -\frac{1}{H_t}\left\{ \mE_t\left[  H_T \int_t^T\frac{\kappa(s)}{\gamma}\md W^\Si_s \right] +  \mE_t\left[ H_T \int_t^T\phi_s\md W^\Oi_s  \right]   \right\}.
\end{align*}
Because
\[
H_s-H_t =-\int_t^s H_r \kappa(r)\md W^{\Si}_r,
\]
we have
\begin{align*}
	&\frac{1}{H_t}\mE_t\left[ H_T \int_t^T\frac{\kappa(s)}{\gamma}\md W^\Si_s  \right]=-\frac{1}{H_t}\mE_t\left[ \int_t^T\frac{|\kappa(s)|^2H_s}{\gamma}\md s\right]=-\mE^{\mQ}_t\left[ \int_t^T\frac{|\kappa(s)|^2}{\gamma}\md s \right],\\
	&\frac{1}{H_t}\mE_t\left[  H_T \int_t^T\phi_s\md W^\Oi_s  \right]=0,
\end{align*}
where $\mQ$ is the risk-neutral measure induced by $H$. To derive the second identity, we have used the fact that $W^\Si$ and $W^\Oi$ are independent. Then
\begin{equation}\label{MVwealth}
	\barX_t=\int_0^t \frac{\kappa(s)}{\gamma}\md W^\Si_s +\int_0^t \phi_s \md W^\Oi_s-\mE^{\mQ}_t\left[ \int_t^T\frac{|\kappa(s)|^2}{\gamma}\md s \right].
\end{equation}
Let us assume now that there exists a function $f(t,s,y)$ which is smooth enough\footnote{This is also assumed in \cite{Basak2010}, and it can be verified directly in specific models. Moreover, $f$ can be obtained by solving a PDE resulting from the Feynman-Kac formula.} such that 
\[
f(t,S_t,Y_t)=\mE^{\mQ}_t\left[  \int_t^T\frac{|\kappa(s)|^2}{\gamma}\md s\right].
\]
Differentiating on both sides of (\ref{MVwealth}) and comparing it to (2.4) yield
\begin{equation}\label{MVstrategy}
	\barpi_t=\frac{\kappa(t)}{\gamma \sigma(t)}-\left(S_t\partial_sf_t+\frac{\rho \nu_t}{\sigma_t}\partial_y f_t \right),
\end{equation}
where $\partial_j f_t=\partial_jf(t,S_t,Y_t)$, $j=s,y$, and $\nu_t=\nu(t,Y_t)$. In addition, as a byproduct, we have
\[
\phi_t=\sqrt{1-\rho^2}\partial_yf_t \nu_t.
\]

To verify that the strategy given by (\ref{MVstrategy}) is indeed an equilibrium, we only  need to check Conditions (a)-(c) in Theorem 3.2. In fact, because the structure of the MV preference is rather simple, all conditions are straightforward except for (3.1), (3.4) and (3.5). Indeed, because $g(\mu)=\int_\mR\left(x-\frac{\gamma}{2}x^2\right)\mu(\md x)+\frac{\gamma}{2}\left(\int_\mR x\mu(\md x)\right)^2$ and $\nabla g(\mu,x)=x-\frac{\gamma}{2}x^2+\gamma\left(\int_\mR y\mu(\md y)\right)x$, we have
\begin{align*}
	g(\mu_1)-g(\mu_0)&=\int_\mR \left(x-\frac{\gamma}{2}x^2\right)(\mu_1-\mu_0)(\md x)+\frac{\gamma}{2}\left(\int_\mR x(\mu_1-\mu_0)(\md x)\right)\left(\int_\mR y(\mu_1+\mu_0)(\md y)    \right) \\
	&=\int_\mR \left[x-\frac{\gamma}{2}x^2+\gamma \left( \int_\mR y\mu_0(\md y) \right)\right](\mu_1-\mu_0)(\md x)+\frac{\gamma}{2}\left(\int_\mR x(\mu_1-\mu_0)(\md x)    \right)^2\\
	&=\int_\mR \nabla g(\mu_0,x)\mu_0(\md x)+M_0(\mu_0,\mu_1),
\end{align*}
with
\[
M_0(\mu_0,\mu_1)\defeq \frac{\gamma}{2}\left(\int_\mR x(\mu_1-\mu_0)(\md x)    \right)^2.
\]
Therefore, (3.1) holds with $M_1\equiv 1$ and $M_0$ as above. Furthermore, 
\begin{align*}
M_0(\mP^{t}_{\barX_T},\mP^{t}_{\barX^{t,\epsilon,\varphi}_T})&=\frac{\gamma}{2}(\mE_t[\barX^{t,\epsilon,\varphi}-\barX_T])^2\\
&=\|\varphi\|^2\frac{\gamma}{2}\left(\mE\left[\int_t^{t+\epsilon}(\theta(s)\md s +\sigma(s)\md W_s^\Si)  \right]     \right)^2\\
&=\|\varphi\|^2\frac{\gamma}{2}\left(\mE\left[\int_t^{t+\epsilon}\theta(s)\md s\right] \right)^2\\
&=O(\epsilon^2).
\end{align*}
Therefore (3.4) and (3.5) are all verified. As such, Theorem 3.2 is applicable to prove that the strategy given by (\ref{MVstrategy}) is an equilibrium. Thus, we obtain the same results as in \cite{Basak2010}. %{\cred However, in this paper the problem has been reformulated (???), in a rigorous and completely different way, and the derivation of the equilibrium strategy is simpler (???)}. 

\section*{OC2 Rank dependent utility.}
Time-consistent portfolio selection with RDU has been studied in \cite{Hu2021}, assuming that the market is complete and both $\theta$ and $\sigma$ are deterministic. We shall recover their ODE (5) using  Condition (a) of our Theorem 3.2. Here, the distortion functions are time-dependent. Hence,  we need a slight extension of the results in the main context. In fact, such an extension is straightforward. Furthermore, we emphasize that, because the preference is now generally non-concave, Theorem 3.2 can not be applied to verify that the solution of the ODE gives an equilibrium. Nevertheless, it can be used to find the candidate equilibrium strategy.  The rigorous verification has been provided in \cite{Hu2021}, which is rather delicate and technical. 

The RDU at time $t$ is given by \[
	g(t,\mP_X)=\int_0^\infty w(t,\mP(U(X)>y))\md y+\int_{-\infty}^0 \left[w(t,\mP(U(X)>y))-1 \right]\md y,
	\]
	where $U$ is the utility function and $w$ is the distortion function.

To proceed, 
%show how Condition (a) of our Theorem 3.2 provides a theoretical back-up for concrete results in \cite{Hu2021},
we need the following result about the derivatives:
%\settheoremtag{OC.1}
%\begin{proposition}\label{RDUTderivative}
%	{\cred If
%	
%	then under some conditions on $w$ and $U$,} we have
	\begin{equation}\label{eq:grad:g:rdu}
	\nabla g(t,\mP_X,x)=\int_{-\infty}^{U(x)}w_p'(t,\mP(U(X)>y))\md y
	\end{equation}
and
	\begin{equation}\label{eq:pd:g:rdu}
	\partial_x \nabla g(t,\mP_X,x)=w'_p(t,1-F_X(x))U'(x),
	\end{equation}
	where $F_X(x)=\mP(X\leq x)$, $x\in\mR$.
%\end{proposition}
%\begin{proof}
	Indeed, suppose that we have imposed enough assumptions on $w$ and $U$ such that we can interchange the differentials and integrals freely. For $\mu_0$ and $\mu_1$, denote $\mu_s=s\mu_1+(1-s)\mu_0$, $s\in[0,1]$. For simplicity, we  use the notation $p^\mu(y)\defeq  \mu(x: U(x)>y) $. Clearly,
	\[
	g(t,\mu_s)=\int_0^\infty w(t,sp^{\mu_1}(y)+(1-s)p^{\mu_0}(y))\md y+\int_{-\infty}^0 [w(t,sp^{\mu_1}(y)+(1-s)p^{\mu_0}(y))-1]\md y.
	\]
	Therefore,
	\begin{align*}
		\frac{\md}{\md s}g(t,\mu_s)&=\int_{-\infty}^\infty w_p'(t,p^{\mu_s}(y))(p^{\mu_1}(y)-p^{\mu_0}(y))\md y\\
		&=\int_{-\infty}^{\infty}w_p'(t,p^{\mu_s}(y))\int_{-\infty}^\infty \ind_{\{U(x)>y\}}(\mu_1-\mu_0)(\md x)\md y\\
		&=\int_\mR\left(\int_{-\infty}^{U(x)}w_p'(t,p^{\mu_s}(y))\md y              \right)(\mu_1-\mu_0)(\md x).
	\end{align*}
By Definition 3.1, we obtain \eqref{eq:grad:g:rdu}. 
%the form of $\nabla g$. 
Taking derivative with respect to the variable $x$ obtains
\eqref{eq:pd:g:rdu}. 
%$\partial_x \nabla g$.
%\end{proof}

Now we are going to derive ODE (5) in \cite{Hu2021}. Following \cite{Hu2021}, we make an ansatz $\barX_T=(U')^{-1}(\nu \cZ^W_T(-\lambda \kappa))$, where $\nu>0$ is a parameter related to the initial endowment, and $\lambda$ is a deterministic function to be determined. For simplicity, we use the following notations:
\begin{align*}
	&\Lambda(t)=\int_t^T |\lambda(s)\kappa(s)|^2\md s,\ \E_{s,t}=\int_s^t \lambda(s)\kappa(s)\md W_s,\ G(x)=-\frac{1}{2}\Lambda(0)+\log \nu-\log U'(x).
\end{align*}
We first calculate $\partial_x \nabla g(t,\mP^{t}_{\barX_T},x)$. As  $\E_{t,T}\sim \N(0,\Lambda(t))$ and is independent of $\F_t$, we have
\begin{align*}
	1-F^{\mP^{t}_{\barX_T}}(x)=\mP_t(\barX_T>x)=\mP_t(G(x)-\E_{0,t}<\E_{t,T})=N\left(\frac{\E_{0,t}-G(x)}{\sqrt{\Lambda(t)}}\right),
\end{align*}
 where we require $\kappa$ to be deterministic,  $N$ is the cumulative distribution function of  the standard normal distribution  $\N(0,1)$. To use Theorem 3.2, we take $\xi^t=\partial_x \nabla g(t,\mP^{t}_{\barX_T},\barX_T)$. 
 
 By \eqref{eq:grad:g:rdu} and \eqref{eq:pd:g:rdu},    we have, for $s\geq t$,
\begin{equation}\label{RDUTeq1}
	\begin{aligned}
		\mE_s[\xi^t]&=\mE_s\left[w'_p\left(t,N\left(\frac{\E_{0,t}-G(\barX_T)}{\sqrt{\Lambda(t)}}\right)\right)\nu\E_T(\lambda\kappa)\right]\\
		&=\mE_s\left[w'_p\left(t,N\left(\frac{-\E_{t,T}}{\sqrt{\Lambda(t)}}\right)\right)\nu e^{-\E_{0,T}+\frac{1}{2}\Lambda(0)}\right]\\
		&=\nu e^{-\E_{0,s}+\frac{1}{2}\Lambda(0)}\mE\left[w_p'\left(t,N\left(\frac{-\E_{t,s}-\sqrt{\Lambda(s)}\xi}{\sqrt{\Lambda(t)}}   \right)\right)e^{\sqrt{\Lambda(s)}\xi}     \right],
	\end{aligned}
\end{equation}
where $\xi\sim \N(0,1)$. Therefore, by It$\hato$'s formula, we have
\begin{align*}
	Z^{\xi^t}(t)&=-\nu e^{-\frac{1}{2}\Lambda(0)-\E_{0,t}}\lambda(t)\kappa(t)\mE\left[w''_p(t,N(\xi))N'(\xi)\frac{e^{\sqrt{\Lambda(t)}\xi}}{\sqrt{\Lambda(t)}}+w'_p(t,N(\xi))e^{\sqrt{\Lambda(t)}\xi}     \right]\\
	&=-\nu e^{-\frac{1}{2}\Lambda(0)-\E_{0,t}}\lambda(t)\kappa(t)\int_\mR\frac{\md }{\md z}\left( w'_p(t,N(z))e^{\sqrt{\Lambda(t)}z}\right)\frac{N'(z)}{\sqrt{\Lambda(t)}}\md z.
\end{align*}
Using integration by part and the fact that $N''(z)=-zN'(z)$ yields
\begin{equation}\label{RDUTeq2}
	Z^{\xi^t}(t)=-\nu e^{-\frac{1}{2}\Lambda(0)-\E_{0,t}}\lambda(t)\kappa(t)\frac{\mE\left[w’_p(t,N(\xi))\xi e^{\sqrt{\Lambda(t)}\xi}\right]}{\sqrt{\Lambda(t)}}.
\end{equation}
Denote $h(t,x)=\mE[w'_p(t,N(\xi))e^{x\xi}]$. Combing (\ref{RDUTeq1}) and (\ref{RDUTeq2}), the equilibrium condition (3.8) becomes
\[
\lambda(t)h'_x(t,\sqrt{\Lambda(t)})=\sqrt{\Lambda(t)}h(t,\sqrt{\Lambda(t)}).
\]
This is in fact an ODE about $\Lambda$. Therefore, the ODE (5) in \cite{Hu2021} is recovered. 

Because the problem with RDU is non-concave, the second-order condition is needed (see inequality (8) in \cite{Hu2021}). We believe that this can be derived by calculating the second-order derivative $\partial^2_x \nabla g(t,\mu,x)$. {To completely accommodate the RDU, it requires a verification theorem for non-concave preferences, which is beyond the scope of this paper and left for the future research.}

\end{document}